\documentclass[%
 reprint,
 superscriptaddress,
 amsmath,amssymb,
 aps,
]{revtex4-2}

\usepackage{graphicx}% Include figure files
\usepackage{dcolumn}% Align table columns on decimal point
\usepackage{bm}% bold math
\usepackage{hyperref}% add hypertext capabilities
\usepackage{placeins}

\begin{document}

\preprint{APS/123-QED}

\title{Overcoming the Curse of Dimensionality: Structural Connectivity Reconstruction via Pairwise Information Flow in Nonlinear Networks}% Force line breaks with \\

\author{Kai Chen}%\email{kchen513@alumni.sjtu.edu.cn}
\altaffiliation{These authors contributed equally to this work.}
\affiliation{School of Mathematical Sciences, Shanghai Jiao Tong University, Shanghai, 200240, China}
\affiliation{Institute of Natural Sciences, Shanghai Jiao Tong University, Shanghai, 200240, China}
\affiliation{Ministry of Education Key Laboratory of Scientific and Engineering Computing, Shanghai Jiao Tong University, Shanghai, 200240, China}
\affiliation{Shanghai Frontier Science Center of Modern Analysis, Shanghai Jiao Tong University, Shanghai, 200240, China}

\author{Zhong-qi K. Tian}%\email{tianzhongqi@alumni.sjtu.edu.cn}
\altaffiliation{These authors contributed equally to this work.}
\affiliation{School of Mathematical Sciences, Shanghai Jiao Tong University, Shanghai, 200240, China}
\affiliation{Institute of Natural Sciences, Shanghai Jiao Tong University, Shanghai, 200240, China}
\affiliation{Ministry of Education Key Laboratory of Scientific and Engineering Computing, Shanghai Jiao Tong University, Shanghai, 200240, China}
\affiliation{Shanghai Frontier Science Center of Modern Analysis, Shanghai Jiao Tong University, Shanghai, 200240, China}

\author{Yifei Chen}%\email{yifeichen99@sjtu.edu.cn}
\affiliation{School of Mathematical Sciences, Shanghai Jiao Tong University, Shanghai, 200240, China}
\affiliation{Institute of Natural Sciences, Shanghai Jiao Tong University, Shanghai, 200240, China}
\affiliation{Ministry of Education Key Laboratory of Scientific and Engineering Computing, Shanghai Jiao Tong University, Shanghai, 200240, China}
\affiliation{Shanghai Frontier Science Center of Modern Analysis, Shanghai Jiao Tong University, Shanghai, 200240, China}

\author{Shouwei Luo}%\email{lsw642878280@sjtu.edu.cn}
\affiliation{School of Mathematical Sciences, Shanghai Jiao Tong University, Shanghai, 200240, China}
\affiliation{Institute of Natural Sciences, Shanghai Jiao Tong University, Shanghai, 200240, China}
\affiliation{Ministry of Education Key Laboratory of Scientific and Engineering Computing, Shanghai Jiao Tong University, Shanghai, 200240, China}
\affiliation{Shanghai Frontier Science Center of Modern Analysis, Shanghai Jiao Tong University, Shanghai, 200240, China}

\author{Songting Li}\email{songting@sjtu.edu.cn}
\affiliation{School of Mathematical Sciences, Shanghai Jiao Tong University, Shanghai, 200240, China}
\affiliation{Institute of Natural Sciences, Shanghai Jiao Tong University, Shanghai, 200240, China}
\affiliation{Ministry of Education Key Laboratory of Scientific and Engineering Computing, Shanghai Jiao Tong University, Shanghai, 200240, China}
\affiliation{Shanghai Frontier Science Center of Modern Analysis, Shanghai Jiao Tong University, Shanghai, 200240, China}

\author{Douglas Zhou}\email{zdz@sjtu.edu.cn}
\affiliation{School of Mathematical Sciences, Shanghai Jiao Tong University, Shanghai, 200240, China}
\affiliation{Institute of Natural Sciences, Shanghai Jiao Tong University, Shanghai, 200240, China}
\affiliation{Ministry of Education Key Laboratory of Scientific and Engineering Computing, Shanghai Jiao Tong University, Shanghai, 200240, China}
\affiliation{Shanghai Frontier Science Center of Modern Analysis, Shanghai Jiao Tong University, Shanghai, 200240, China}
\affiliation{State Key Laboratory of Synergistic Chem-Bio Synthesis, Shanghai Jiao Tong University, Shanghai, 200240, China}

\date{\today}% It is always \today, today,

\begin{abstract}
Inferring structural connectivity from observed dynamics remains a fundamental open problem in complex systems, particularly for nonlinear networks where direct measurements are unavailable, and existing methodological approaches each incur characteristic limitations. Model-based methods require prior knowledge of the mechanistic form of the underlying dynamics, while model-free approaches often lack quantitative correspondence to network structural connectivity, and suffer from the curse of dimensionality as the size and complexity of the system increases. Here we show that pairwise time-delayed information flow is sufficient to recover, without high-dimensional conditioning, structural connectivity in general nonlinear networks. We introduce a pairwise delayed information flow (PDIF) as an information-theoretic framework and derive a theoretical quadratic relationship between PDIF and coupling strength, establishing a direct correspondence between information flow and network architecture. We further show that indirect interaction contributions are suppressed at leading order, enabling accurate reconstruction solely from pairwise measurements. Combining binary state representations, pairwise inference, and time-delayed statistics, PDIF overcomes the dimensionality barrier while remaining model-agnostic and scalable. Validated across nonlinear dynamical systems, neuronal network models, and large-scale electrophysiological recordings, PDIF achieves high reconstruction accuracy and robustness to noise, outperforming existing methods. These results establish a principled, efficient and model-agnostic framework for connectivity reconstruction, and reveal a general mechanism by which pairwise observable statistics encode network structure in nonlinear systems.

% \begin{description}
% Secondary publications and information retrieval purposes.
% You may use the \texttt{description} environment to structure your abstract;
% use the optional argument of the \verb+\item+ command to give the category of each item. 
% \end{description}
\end{abstract}

% \keywords{Suggested keywords}%Use showkeys class option if keyword
\maketitle

%\tableofcontents

\section{\label{sec:intro}Introduction}
The structure of complex networks governs their dynamics across scientific domains, including neuroscience \cite{suarez2020linking} and climatology \cite{boers2021complex}. Yet direct measurement of structural connectivity is often technologically constrained, motivating a broad spectrum of methodological approaches that seek to reconstruct connectivity from observed node activity time series. When the mechanistic properties of the system are sufficiently well characterised, model-based approaches — including sparse regression \cite{napoletani2008reconstructing}, compressed sensing \cite{barranca2023reconstruction}, and equation-learning methods \cite{brunton2016discovering,casadiego2017modelfree} — exploit structural sparsity or libraries of candidate interaction terms to recover network topology from time series. While these methods afford interpretability by enabling direct mechanistic insights into coupling structure, they incur substantial limitations: they require governing equations to be expressible within a predefined functional basis, rendering inference systematically unreliable when the true dynamics fall outside this constraint, and they demand high-quality time series of sufficient volume, a requirement that becomes increasingly stringent as network complexity grows. Model-free statistical approaches, by contrast, avoid mechanistic assumptions by operating directly on observable statistics. Correlation-based methods \cite{bedenbaugh1997multiunit} are computationally tractable but conflate direct and indirect interactions, providing no principled route to structural connectivity. Granger causality (GC) \cite{ge2012componential,zhou2013causal,mehdizadehfar2020brain} captures directed causal dependence through second-order statistics, yet fails to account for higher-order dependencies characteristic of nonlinear systems. Bayesian network methods \cite{friedman2000using,mumford2014bayesian} offer principled uncertainty quantification but become computationally intractable as the graph space grows super-exponentially with node count. Transfer entropy (TE) \cite{schreiber2000measuring} and its conditional variants \cite{shahsavaribaboukani2020estimating,papana2012detection} provide model-free, information-theoretic measures of directed causal influence and have been widely applied in neuroscience, finance, and social science \cite{honey2007network,versteeg2012information,li2013risk}.

Across this methodological diversity, two fundamental limitations recur. The first is the absence of a quantitative, mechanistic correspondence between inferred measures and true coupling strength. In model-based approaches, such a correspondence is valid only when the assumed model or functional basis accurately reflects the underlying dynamics, which is rarely verifiable for complex, model-agnostic systems. Model-free approaches, while avoiding this assumption, typically quantify statistical association between nodes in terms — entropy reduction, posterior edge probability, partial correlation — that do not admit direct translation into coupling strength.
The second limitation is the curse of dimensionality, which manifests across method families in distinct but equally prohibitive forms: sparse regression and equation-learning approaches face a combinatorial search over candidate interaction terms; Bayesian methods confront a super-exponential expansion of the graph structure space; and information-theoretic measures such as TE require estimation of joint distributions over high-dimensional node histories whose data requirements scale exponentially with network size and embedding depth. Resolving both challenges — the absence of a quantitative coupling-strength correspondence and the curse of dimensionality — in a general, model-agnostic setting remains an open problem.

Among these approaches, TE has attracted particular interest as an information-theoretic measure of directed causal influence. For an interaction $X \to Y$, TE quantifies the information flow from $X$ to $Y$ as the reduction in uncertainty about $Y$'s future attributable to knowledge of $X$'s past. Both limitations identified above manifest in TE with particular acuity. On the theoretical side, while GC — a linear counterpart of TE, to which it is equivalent for Gaussian variables \cite{barnett2009granger} — provides a partial foundation by establishing a quantitative link between inferred connectivity and structural connectivity in linear and certain specific nonlinear systems \cite{bressler2011wiener,zhou2013causal}, this result does not extend to the general nonlinear case. On the computational side, conditional TE and its variants \cite{shahsavaribaboukani2020estimating,papana2012detection}, designed to isolate direct from indirect causal effects, compound this burden further by requiring simultaneous observation of all network nodes — a condition rarely satisfied in practice. Prior mitigation strategies — including ordinal symbolisation \cite{staniek2008symbolic}, $k$-nearest neighbour density estimation \cite{gomez-herrero2015assessing}, kernel methods \cite{darmon2017specific}, sparsity exploitation \cite{runge2012escaping}, and first-order Markov approximations \cite{sipahi2020improving} — each address part of the problem but none resolves it in full generality.

Here we resolve both challenges simultaneously by introducing a pairwise delayed information flow (PDIF) framework. PDIF is an information-theoretic measure inspired by TE, yet the results it enables reach well beyond the conventional TE setting. PDIF quantifies directed information flow between pairs of nodes using binary state representations and pairwise time-delayed statistics, eliminating the need for high-dimensional joint distribution estimation. On this basis, we derive a theoretical quadratic relationship between PDIF and inter-node coupling strength, establishing a direct and quantitative correspondence between observable information flow and network architecture in general nonlinear systems. Crucially, we prove that contributions from indirect interactions are suppressed at leading order — a result that reflects a structural property of information flow in nonlinear dynamics — showing that pairwise measurements suffice for accurate reconstruction without conditioning on the full network state. Together, these properties enable PDIF to overcome the dimensionality barrier while remaining model-agnostic and scalable to large networks without additional structural assumptions. Using the Hodgkin-Huxley (HH) neuronal network \cite{hodgkin1952quantitative} as a primary testbed, we demonstrate precise reconstruction from binarized spike trains, and then extend the framework to multi-compartment cerebellar circuits \cite{hull2022cerebellar,dangelo2016modeling}, Lorenz networks \cite{lorenz2017deterministic,anishchenko1998synchronization}, logistic networks \cite{may1976simple,dickten2014identifying}, Rössler networks \cite{rossler1976equation}, and random recurrent neural networks \cite{sompolinsky1988chaos}. PDIF outperforms existing methods across all benchmark systems. Applied to large-scale electrophysiological recordings from the Allen Institute \cite{siegle2021survey,allendata}, PDIF yields consistent and robust reconstructions across brain state conditions. Collectively, these results establish a principled, efficient, and model-agnostic framework for inferring structural connectivity from observed dynamics, and reveal a general mechanism by which pairwise observable statistics encode network structure in nonlinear systems.

\section{\label{sec:results}Results}
\subsection{Pairwise delayed information flow (PDIF)}

Consider a nonlinear network of $s+2$ nodes, 
nodes $X$ and $Y$ are two different nodes in the network, and $Z_i$, ($i=1,2,\cdots,s$) is one of the resting nodes.
$\{x_n\}$, $\{y_n\}$, and $\{z_{i,n}\}$ are the time series of nodes' activity of $X$, $Y$, and $Z_i$, respectively.
If there exists structural connection from $X$ to $Y$, the information contained in $X$'s activity must flow towards that in $Y$. Inspired by this fact, we define the measure of 
PDIF from $X$ to $Y$ as follows: 
\begin{eqnarray}
I_{X\rightarrow Y}(k,l,\tau)=\sum\limits_{y_{n+1},y_{n}^{(k)},x_{n-\tau}^{(l)}} p\left(y_{n+1},y_{n}^{(k)},x_{n-\tau}^{(l)}\right)\nonumber\\
\cdot\log\frac{p\left(y_{n+1}|y_{n}^{(k)},x_{n-\tau}^{(l)}\right)}{p\left(y_{n+1}|y_{n}^{(k)}\right)},
\label{eq:TE}
\end{eqnarray}
where $\tau$ is the delay parameter considering the latency of information flow.
$y_{n+1}$ is the future state of $Y$, and
$y_{n}^{(k)}:=[y_{n},y_{n-1},...,y_{n-k+1}]$ and
$x_{n-\tau}^{(l)}:=[x_{n-\tau},x_{n-\tau-1},...,x_{n-\tau-l+1}]$
are the past state of $Y$ and $X$, respectively.
$k$ and $l$ are the order parameters for $Y$ and $X$, respectively, characterizing the length of past history used in estimating the information flow.

Conceptually, PDIF is an information-theoretic measure which is related to TE and conditional TE.
Unlike these conventional statistical measures, PDIF overcomes the curse of dimensionality and provides a scalable framework for flexibly probing the information flow in a pairwise manner, mechanistically capturing the underlying structural connectivity through
three key aspects.
First, rather than being applied directly to continuous-valued time series,
PDIF incorporates a binarization preprocessing step
inspired from action potential generation and
spike-train recordings widely analyzed in the neuroscience.
This enables the use of a smaller order parameter $k$, significantly 
reducing the number of states in the joint probability state space (the detailed rationale is discussed later).
Second, PDIF adopts a pairwise analysis framework, ensuring broader
applicability to large networks with many nodes.
In practice, partial observations often afford 
access only to subnetworks.
Unlike conventional TE-related methods, which require estimation of high-dimensional joint probability distribution
$p\left(y_{n+1},y_{n}^{(k)},x_{n}^{(l)}, z_{1,n}^{(k_1)},\cdots,z_{s,n}^{(k_s)}\right)$ (where 
$z_{i,n}^{(k_i)}:=[z_{i,n},z_{i,n-1},...,z_{i,n-k_{i}+1}]$),
PDIF operates independently of third-party node activity, enabling reconstruction
of subnetworks even when $\{z_{i,n}\}$ ($i=1,\cdots,s$)
are not fully available. This bypasses complications arising from hidden nodes and
substantially reduces the dimensionality of the joint probability space.
Finally, motivated by the concept of information-transfer delays \cite{wibral2013measuring}, PDIF introduces an explicit delay parameter $\tau$ to further
reduce the requisite order parameter $l$ in practical applications. 

\subsection{Network reconstruction pipeline using PDIF}

\begin{figure*}[htp]
    \centering
    \includegraphics[width=\textwidth]{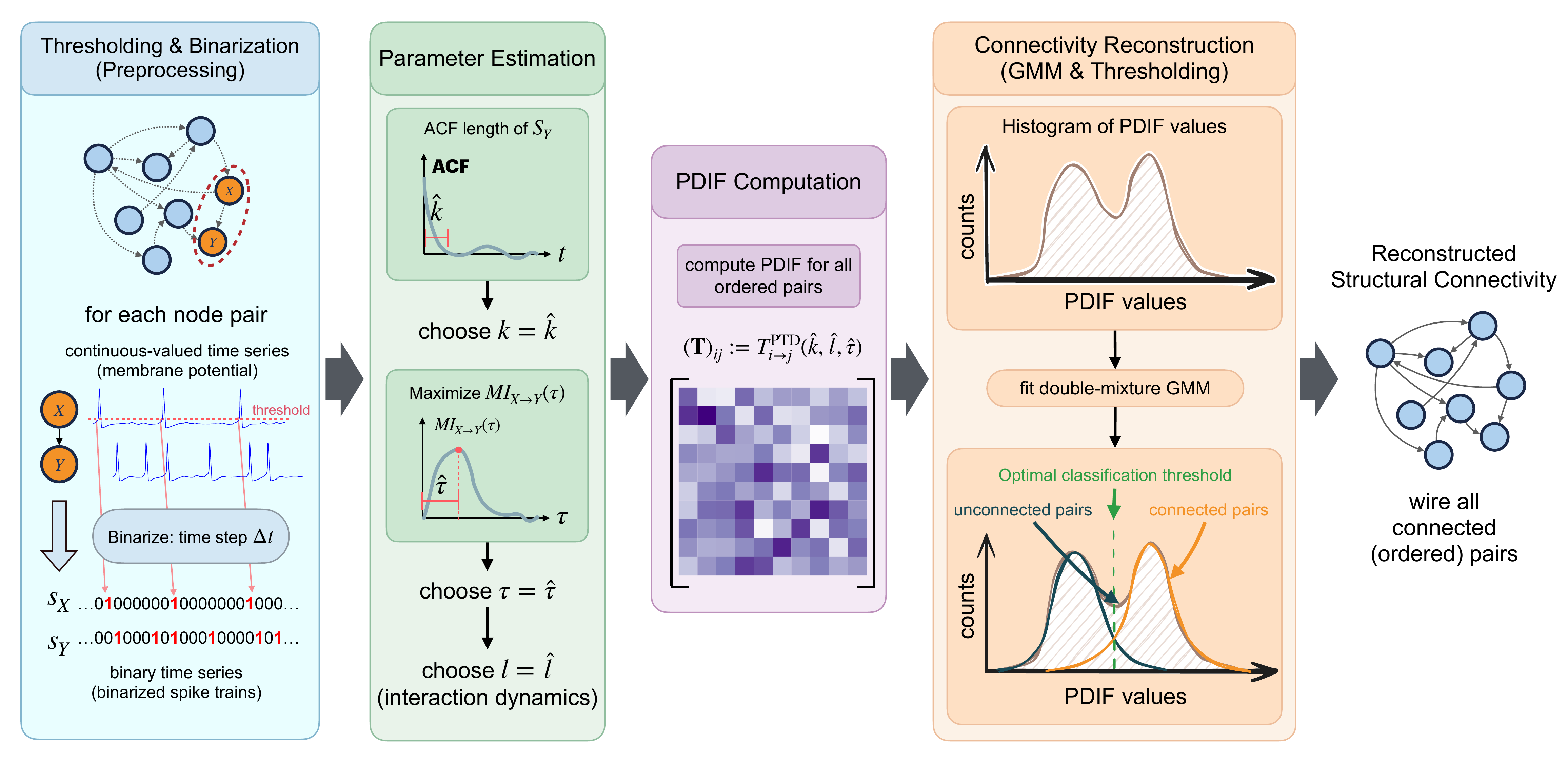}
    \caption{\textbf{Schematics of PDIF reconstruction pipeline.}}
    \label{fig:recon-pipeline}
\end{figure*}

We develop a framework to reconstruct the structural
connectivity ({i.e.}, the network's adjacency matrix) using 
PDIF-estimated effective connectivity. This involves solving a binary
classification problem to distinguish directly connected node pairs.
First, we compute PDIF values for all possible node pairs in a network.
For a given ordered pair, say $X\to Y$, we determine PDIF parameters $k$, $l$ and $\tau$
through the following pipeline, as shown in Fig. \ref{fig:recon-pipeline}:
(i) convert continuous-valued time series $\{x_n\}$ and $\{y_n\}$ to spike trains by
thresholding, and then binarize them using discrete time step $\Delta t$,
(ii) compute the autocorrelation function (ACF) of the binarized $\{y_n\}$ to select $k=\hat{k}$,
(iii) estimate the optimal delay $\tau=\hat{\tau}$ using
$\hat\tau = \arg\max_{\tau}MI_{X\to Y}(\tau)$ (where $MI_{X\to Y}(\tau)$ is the time-delayed mutual information from $X$ to $Y$ given delay $\tau$), and
(iv) choose $l$ based on interaction dynamics, e.g., $l=1$ for most cases; higher for weak and persistent interactions (see
more details in Methods).

Next, we fit a double-component Gaussian mixture model (GMM) to the distribution of PDIF values across all pairs.
The binary classification boundary is determined by an optimal threshold to distinguish two Gaussian mixtures (green dashed line in Fig.~\ref{fig:recon-pipeline}), 
identifying directly connected pairs as those with PDIF values above
the threshold.

\subsection{Reconstructing Hodgkin-Huxley neuronal networks}
We test the effectiveness of PDIF-based network reconstruction using the classical
Hodgkin-Huxley (HH) neuronal network model \cite{hodgkin1952quantitative}, 
investigating the relationship between PDIF-inferred effective connectivity and structural connectivity.
The HH model is one of the most widely used biophysical neuronal models, which quantitatively
describes the ionic mechanisms of membrane potential dynamics and action potential generations. The membrane dynamics, $V_i(t)$, is defined by
\begin{eqnarray}
C\dot{V}_i = I_\mathrm{L}+I_\mathrm{Na}+I_\mathrm{K}
-(V_{i}-V_\mathrm{F})G_{i}^{\text{F}} \nonumber\\
- (V_{i}-V_\mathrm{E}) G_{i}^{\text{E}}
- (V_{i}-V_\mathrm{I}) G_{i}^{\text{I}},
\label{eq:HH-simplified}
\end{eqnarray}
where $C$ is the membrane capacitance. $I_\mathrm{L},I_\mathrm{Na}$, and $I_\mathrm{K}$
are the leaky, sodium, and potassium currents, respectively.
The nonlinearity of sodium and potassium channels is depicted by high-order nonlinear
gating dynamics (see details in Methods).
The last three terms in Eq. (\ref{eq:HH-simplified}) represent the synaptic inputs from
external Poisson drives, excitatory and inhibitory recurrent interactions from other neurons,
respectively. $V_\mathrm{F}$, $V_\mathrm{E}$, and $V_\mathrm{I}$ are their corresponding
reversal potentials. Their corresponding conductances, $G_i^\mathrm{F}$, $G_i^\mathrm{E}$,
and $G_i^\mathrm{I}$, are defined by
\begin{eqnarray*}
G_{i}^{\text{F}}(t)&&=f\sum_{k}H(t-T_{ik}^{\text{F}}), \\
G_{i}^{\text{E}}(t)&&=\sum_{j}^{N_E}S_{ij}\sum_{k}H(t-T_{jk}),\\
G_{i}^{\text{I}}(t)&&=\sum_{j}^{N_I}S_{ij}\sum_{k}H(t-T_{jk}),
\end{eqnarray*}
where $f$ is the strength of the Poisson process with rate $\nu$,
and $S_{ij}$ is the coupling strength from $j$-th neuron to $i$-th neuron.
$T_{ik}^{\text{F}}$ is the arrival time of the $k$-th Poisson spike for 
the $i$-th neuron, and $T_{jk}$ is the arrival time of the $k$-th recurrent
synaptic input from $j$-th neuron to $i$-th neuron.
The temporal course of conductance change induced by
a single Poisson or recurrent spike is modeled as a double exponential function
$H(t)=\frac{\sigma_{r}\sigma_{d}}{\sigma_{d}-\sigma_{r}}[\text{exp}(-t/\sigma_{d})-\text{exp}(-t/\sigma_{r})]\Theta(t)$,
where $\sigma_{r}$ and $\sigma_{d}$ are  the rise and decay timescales, 
respectively, and $\Theta(t)$ is the Heaviside function.
When the voltage $V_{i}$, exceeds the threshold $V^{\textrm{th}}$ at $T_{ik}$,
we denote this event as the $k$-th spike of the $i$-th neuron. This triggers synaptic inputs to all its downstream neurons.  More details about HH model and its parameters can be found in Methods. By analyzing HH neuronal networks, we can validate PDIF's potential applicability to real neural recordings.

\begin{figure*}[htp]
    \centering
	\includegraphics[width=\textwidth]{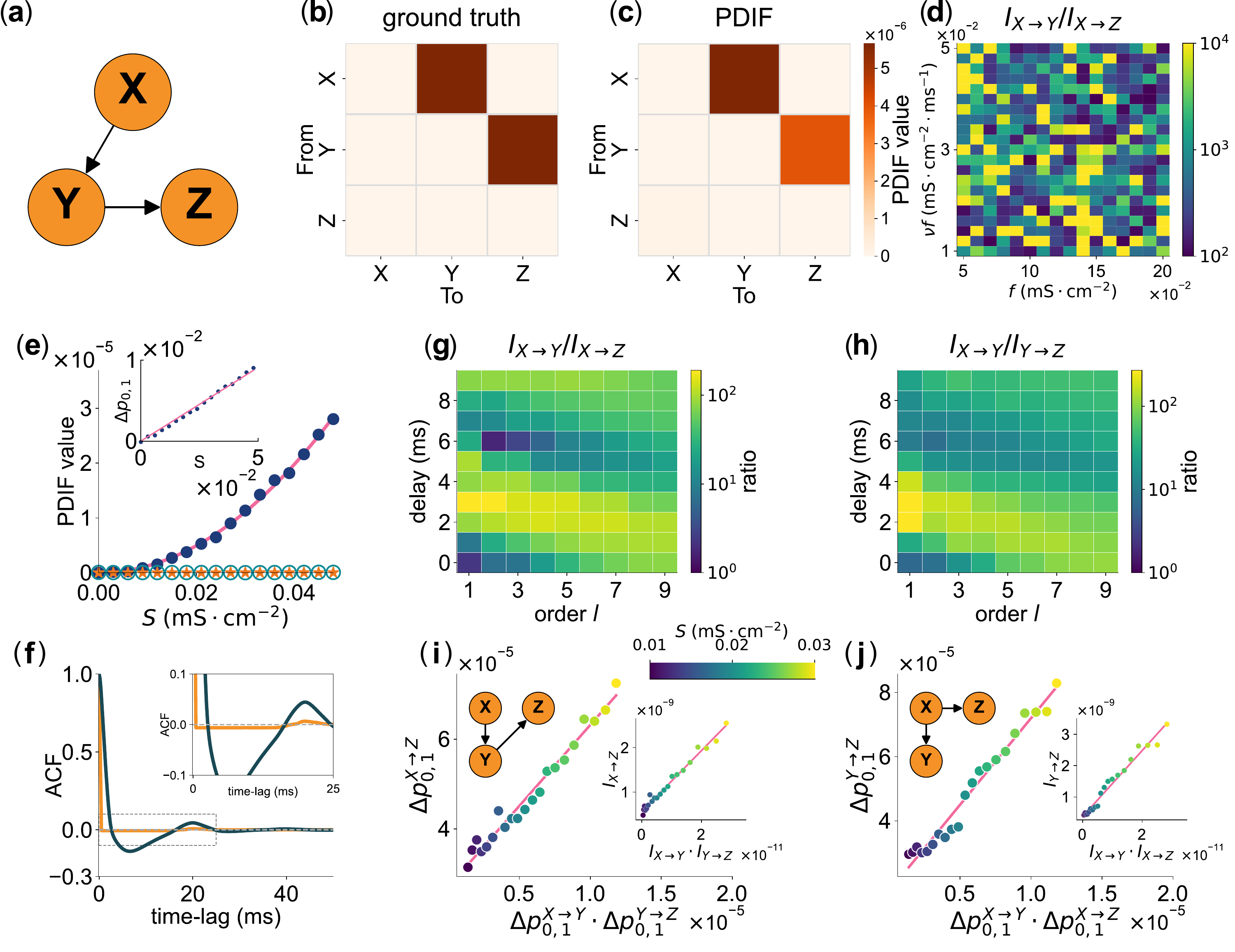}
	\caption{\textbf{PDIF reconstruction performance of three-neuron HH networks and corresponding underlying reconstruction mechanisms.}
    (\textbf{a}) The graph of a three-neuron network with chain motif.
    (\textbf{b}) The adjacency matrix for the network in (\textbf{a}).
    (\textbf{c}) The matrix of inferred PDIF values for the network in (\textbf{a}). 
    (\textbf{d}) The ratio of PDIF values between connected and unconnected pairs, $I_{X\to Y}/I_{X\to Z}$. 
	The minimum ratio in (\textbf{d}) is greater than 100,
    across a broad dynamical regime with firing rates from $2$ to $50$ Hz.
    (\textbf{e}) PDIF values versus coupling strength $S$ for the HH
    network in (\textbf{a}).
    The blue dots, orange stars, and green circles indicate
    $I_{X\rightarrow Y}$, $I_{Y\rightarrow X}$,
    and $I_{X\rightarrow Z}$, respectively.	
    The pink line indicates a quadratic fit.
    Inset: linear relation between $\Delta p_{0,1}$ and $S$ from $X$ to $Y$ with
    the pink line reflecting a linear fit.
    (\textbf{f}) ACF of (orange) binary spike-train data
    and (dark green) voltage time series of neuron $X$.
    Inset: the zoom of the region inside the dashed square in ACF plot.
    (\textbf{g})-(\textbf{h})
    The ratio between PDIF values of connected and unconnected
    neuronal pairs (color coded) as varying order $l$ and delay $\tau$
    in the three-neuron HH system with chain motif (see (\textbf{a})) and confounder motif (see Fig. \ref{fig:HH3-confounder}(\textbf{a})), respectively.
    (\textbf{i})-(\textbf{j}) 
    The relation of $\Delta p_{a,b}$ values between connected and unconnected pairs
    in the three-neuron HH system with chain motif (see \textbf{a}) and confounder motif (see Fig. \ref{fig:HH3-confounder}(\textbf{a})), respectively.
    Inset: the relation between the corresponding PDIF values for those two types of
    neuronal pairs. The color of scatter plots indicates the homogeneous coupling strength
    $S$ in networks.
    Here, PDIF values are estimated from $2.7$ hours of neuronal recordings. 
    Unless explicit stated, the parameters are chosen as $k=1$, $l=1$, and $\tau=3$ ms, following the pipeline in Fig. \ref{fig:recon-pipeline}.
    }\label{fig:HH3}
\end{figure*}

We first build a three-neuron HH network connected with the chain motif, as shown in Fig. \ref{fig:HH3}(a)-(b),
simulating it using a fourth-order Runge-Kutta scheme. The spike times of neurons, $\{T_{ik}\}\,(i=1,2,3)$, are recorded
(see HH model details in Methods).
With a proper time step $\Delta t = 0.5\,\mathrm{ms}$, we convert the spike trains into binary time series $S_i(t)$, where $S_i(t)=1$ indicates that $i$-th neuron fires a spike within $[t,t+\Delta t]$, and $S_i(t)=0$ otherwise.
We emphasize that the spiking nature of neurons and synaptic interactions enables this
continuous-to-binary conversion of neurons' voltages,
which inspired the binarization step in the PDIF-reconstruction pipeline originally.
We later demonstrate its extension to non-neuronal network systems.

Following the PDIF reconstruction pipeline (Fig. \ref{fig:recon-pipeline}), the estimated PDIF values 
align closely with the structural connectivity pattern, i.e., the adjacency matrix (Fig.
\ref{fig:HH3}(c)).
Notably, PDIF values for connected neuron pairs ($X\to Y$ or $Y\to Z$)
are significantly larger - often by orders of magnitude - than those for unconnected pairs ($X\to Z$).
To assess robustness, we simulate multiple 
realizations of the three-neuron system by varying Poisson input parameters (input strength $f$ and input frequency $\nu$), which primarily control the system's dynamical regime.
We then evaluate the ratios $I_{X\to Y}/I_{X\to Z}$.
As shown in Fig. \ref{fig:HH3}(d), this ratio consistently exceeds 
$100$ across diverse dynamical regimes, {i.e.}, $I_{X\to Y}$ for the connected pair is at least 2 orders of magnitude larger than $I_{X\to Z}$ for the unconnected pair.
This robust performance of PDIF effectively distinguishes connected pairs from those indirectly interacting ones, ensuring reliable network reconstruction.
We further evaluate PDIF on a three-neuron HH network with a confounder motif (Fig. \ref{fig:HH3-confounder}), where one neuron influences two others without direct interaction between them. Consistent with previous results, PDIF effectively distinguishes direct connections from indirect ones, as demonstrated in Fig. \ref{fig:HH3-confounder}.

\subsection{Mechanisms underlying PDIF reconstruction}

The remarkable effectiveness of PDIF reconstruction motivates us to explore
its underlying mechanism.
First, the consistent alignment between PDIF values and structural connectivity suggests a quantitative relationship between them, which we explore further.
Second, unlike conventional TE-related statistical measures, PDIF avoids the curse of dimensionality, which typically arises when applied to continuous-valued time series. 

\subsubsection{PDIF correlates with network coupling strength}

We next derive the relationship between PDIF and network coupling strength by considering a two-neuron network with a unidirectional connection ($X\rightarrow Y$). To simplify the notation, we first map each binary number sequence of $y_{n}^{(k)}$ and $x_{n-\tau}^{(l)}$ in Eq. (\ref{eq:TE}) to a decimal number ({e.g.}, the state [1,0,0] is mapped to 4 as decimal number representation). And for simplicity, we denote  
\begin{eqnarray*}
p_{a,b}=&&p\left(y_{n+1}=1|y_{n}^{(k)}=a,x_{n-\tau}^{(l)}=b\right)\\
\Delta p_{a,b} =&&p_{a,b}-p_{a,0},
\end{eqnarray*}
where $a$ and $b$ are decimal numbers representing the state of $y_{n}^{(k)}$ and $x_{n-\tau}^{(l)}$, respectively.
Note that $\Delta p_{a,b}$ reflects the activity induced change of state probability of $y_{n+1}=1$ given the non-quiescent $x_{n-\tau}^{(l)}$ state, which depends on coupling strength $S$.
Subsequently, we can rewrite PDIF expression (Eq. (\ref{eq:TE}))
in terms of $p_{a,b}$ and $\Delta p_{a,b}$ with a Taylor expansion form:
\begin{eqnarray}
I_{X\rightarrow Y}= \frac{1}{2}\sum_{a}\frac{1}{p_{a,0}-p_{a,0}^{2}}\left\{\sum_{b}p(a,b)\Delta p_{a,b}^{2}\right. \nonumber\\
\left.-\frac{\left[\sum_{b}p(a,b)\Delta p_{a,b}\right]^{2}}{p(a)}\right\}+o\left(\Delta p_{a,b}^{2}\right),
\label{eq:TE_dp}
\end{eqnarray}
where $p(a,b)=p(y_{n}^{(k)}=a,x_{n-\tau}^{(l)}=b)$
and $p(a)=p(y_{n}^{(k)}=a)$
(see Methods for derivation details).
From Eq. (\ref{eq:TE_dp}), we reveal that the PDIF value is quadratically
related to $\Delta p_{a,b}$.
We then show that $\Delta p_{a,b}$ can be Taylor expanded with respect to coupling strength $S$ 
(see Methods for derivations), with a non-vanishing first-order leading term, {i.e.}, $\Delta p_{a,b} \propto S$ to the leading order.
We numerically verify the linear relation by systematically simulating
an ensemble of a three-neuron network (shown in Fig. \ref{fig:HH3}(a))
with various values of $S$ and estimating the corresponding value of $\Delta p_{a,b}$,
as shown in the inset of Fig. \ref{fig:HH3}(e). 
Consequently, PDIF is proportional to $S^2$, with the numerical verification shown in Fig. \ref{fig:HH3}(e), 
which explains why the structural connectivity pattern in HH
networks can be accurately inferred by PDIF values.

\subsubsection{PDIF overcomes the curse of dimensionality}

Next, we again employ the same $(s+2)$-node network to illustrate 
how PDIF overcomes the dimensionality problem, faced by other information-theoretic measures.
To estimate an information-theoretic measure from $X$ to $Y$,
conventional approaches require estimation of
the high-dimensional joint probability distribution $p\left(y_{n+1},y_{n}^{(k)},x_{n}^{(l)}, z_{1,n}^{(k_1)},\cdots,z_{s,n}^{(k_s)}\right)$,
whereupon the number of states in the probability space grows exponentially with the total 
dimensionality $(k + l + 1 + \sum_{i=1}^s k_i)$, with $k_i$ denoting the order parameter for $Z_{i,n}$.
\cite{schreiber2000measuring,staniek2008symbolic}.
Considering continuous-valued time series of node activity, 
if the range of neural voltage is partitioned into $p$ uniform bins for distribution estimation, then total number of states becomes $p^{k+l+1+\sum_{i=1}^s k_i}$,
rendering accurate probability estimation intractable under limited neural recording data --
a manifestation of the curse of dimensionality. In contrast, PDIF successfully addresses this
challenge simultaneously along four dimensions: a reduced bin count $p$, a reduced history order parameter $k$ for $Y$, a reduced history order parameter $l$ for $X$, and the elimination of order parameter $k_i$ (i.e., $k_i=0$). 

First, the binary state of the signals allows us to naturally set $p=2$, which substantially reduces the dimension of the probability space. 

Second, the binary state of the signals permits the use of a small $k$. In conventional TE-based reconstruction, the order $k$ is introduced to prevent the mis-inference due to strong history dependence of a signal \cite{schreiber2000measuring}. In general, $k$ should be chosen at which the signal's auto-correlation function (ACF) decays to nearly zero, as shown in Fig. \ref{fig:recon-pipeline}.
For the binary time series from neuronal spiking activity, the ACF decays to near-zero within $1$ ms (orange in Fig. \ref{fig:HH3}(f)),
which is significantly faster than for continuous-valued voltage time series (dark green in Fig. \ref{fig:HH3}(f)).
As a near-zero ACF indicates uncorrelatedness with the signal's own history, allowing the binary time series to be treated as independent random variables (see Ref. \cite{zhang2019bet} that shows uncorrelatedness is equivalent to independence for binary random variables).
This ensures the validity of using a small order parameter $k=\hat{k}$ in PDIF. 
In the remaining part of this work, we choose $\hat{k}=1$ for reconstructing HH networks (with $\hat{k}$ selected similarly for other nonlinear networks
based on their ACFs).

Third, by introducing an additional time delay parameter, PDIF further reduces the order $l$. In conventional TE-based reconstruction, 
$l$ must be chosen to cover the typical time-scale of the inter-neuronal interactions to distinguish connected neuronal pairs from unconnected ones.
However, following the pipeline (Fig. \ref{fig:recon-pipeline}), 
PDIF determines the optimal time delay by scanning across delays 
to maximize the mutual information between neuron pairs,
thereby improving the separability of connected and unconnected pairs with minimal $l$.
Considering the three-neuron system shown in Fig. \ref{fig:HH3}(a),
with the optimal time delay $\tau=3$ ms, PDIF achieves the best reconstruction performance at $l=1$, {i.e.} the ratio
$I_{X\to Y}/I_{X\to Z}$ reaches its maximum.
In contrast, if one simply sets $\tau=0$ ms, it requires $l\ge 5$ for achieving comparable results, as shown in Fig. \ref{fig:HH3}(g).
For the three-neuron system in Fig. \ref{fig:HH3-confounder}(a), we obtain similar results as shown in Fig. \ref{fig:HH3}(i).

Finally, PDIF operates pairwise, eliminating the need to account for the state dimensionality of other neurons in the network. For the three-neuron system illustrated in Fig. \ref{fig:HH3}(a) (\ref{fig:HH3-confounder}(a)),
conventional approaches exclude the indirect causal relation between $X$ and $Z$ ($Y$ and $Z$)
by conditioning on the activity of neuron $Y$ ($X$), thereby introducing extra dimensionality.
In contrast, PDIF inherently resists the influence of indirect causal effects, as evidenced by the accurate reconstructions in Fig. \ref{fig:HH3}(c) and \ref{fig:HH3-confounder}(c).
To understand this, we first analyze the relation between PDIF values for disconnected and connected neuron pairs in a chain motif (Fig. \ref{fig:HH3}(a)).
Without loss of generality, we express $\Delta p_{a,b}^{X\rightarrow Z}$ as a double-variate function $\Delta p_{a,b}^{X\rightarrow Z} =f(\Delta p_{a,b}^{X\rightarrow Y}, \Delta p_{a,b}^{Y\rightarrow Z})$.
By performing Taylor expansion with respect to coupling strengths $S^{X\to Y}$ and $S^{Y\to Z}$ and retaining the dominant terms (see Methods for derivation details),
we derive 
\begin{equation*}
    \Delta p_{a,b}^{X\rightarrow Z}=O(\Delta p_{a,b}^{X\rightarrow Y}\cdot \Delta p_{a,b}^{Y\rightarrow Z}).
\end{equation*}
The numerical verification of this relation is shown in Fig. \ref{fig:HH3}(h).
Notably, both $\Delta p_{a,b}^{X\rightarrow Y}$ and $\Delta p_{a,b}^{Y\rightarrow Z}$ of direct causality are very small (typically $|\Delta p_{a,b}|<0.01$ for $\Delta t = 0.5$ ms calculated using parameters in neurophysiological regime).
Consequently, $\Delta p^{X\rightarrow Z}_{a,b}$ is much smaller than
$\Delta p^{X\rightarrow Y}_{a,b}$ and $\Delta p_{a,b}^{Y\rightarrow Z}$.
Combining with Eq. (\ref{eq:TE_dp}), $I_{X\rightarrow Z}$ (green circles) is negligible compared with $I_{X\rightarrow Y}$ (blue dots) as shown in Fig. \ref{fig:HH3}(e).
Therefore, although we calculate PDIF without conditioning on the information of intermediate nodes, the analysis shows that direct causal interactions can be distinguished from indirect ones.
Similarly, for the confounder motif of $Y\leftarrow X\rightarrow Z$, we derive
\begin{equation*}
\Delta p_{a,b}^{Y\rightarrow Z}=O(\Delta p_{a,b}^{X\rightarrow Y}\cdot \Delta p_{a,b}^{X\rightarrow Z}), 
\end{equation*}
which is numerically verified in Fig. \ref{fig:HH3}(j). Here, $\Delta p_{a,b}^{Y\rightarrow Z}$ is orders of magnitude smaller than $\Delta p_{a,b}^{X\rightarrow Y}$ and $\Delta p_{a,b}^{X\rightarrow Z}$, ensuring that the PDIF values between commonly driven nodes are 
negligible compared to those with direct interactions. 

Overall, to reconstruct a connection from $X$ to $Y$, compared with the state number $p^{k+l+1+\sum_i^sk_i}$ in conventional approaches,
PDIF reconstruction pipeline only requires estimating the joint probability with the state number $2^{\hat{k}+l+1}$ ($2^3$ for HH networks).
This dramatic reduction of required state numbers makes the PDIF reconstruction applicable in practice.

\subsection{Application of PDIF to large neuronal networks}

\begin{figure*}[htp]
    \centering
	\includegraphics[width=\textwidth]{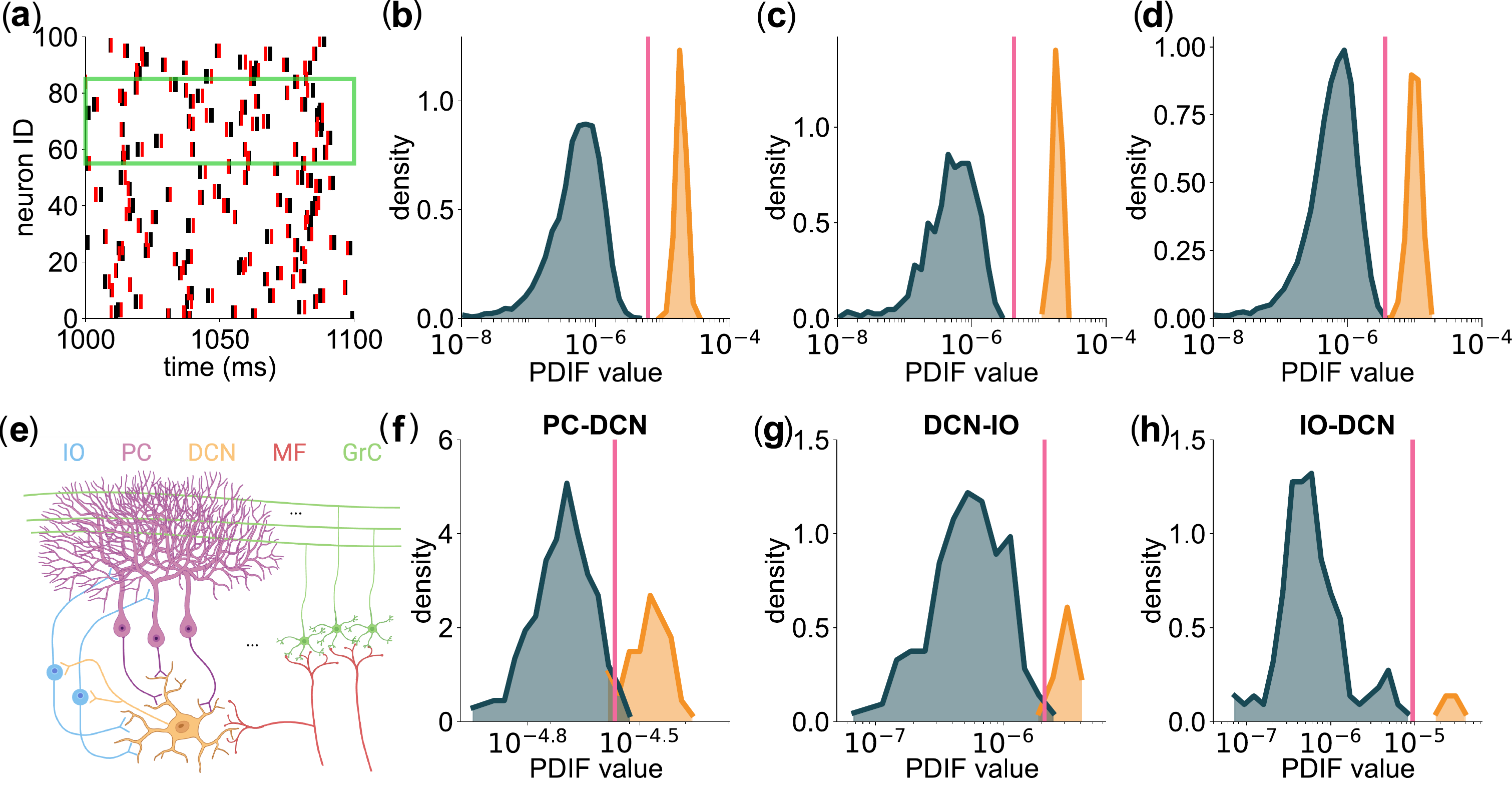}
	\caption{\textbf{PDIF reconstruct large neuronal networks.}
    	(\textbf{a}) The raw (black bars) and the temporally jittered 
        (red bars) raster plots for a 100-neuron HH network. The green
        box indicates the range of neuronal indices for a 30-neuron subnetwork,
        sampled for analysis in (\textbf{c}).
        (\textbf{b}) The performance of PDIF-based reconstruction
        for the HH network in (\textbf{a}) using its raw spike trains (ACC = 100\%, AUC = 1). 
        (\textbf{c}) The performance of PDIF-based reconstruction for 
        a 30-neuron subnetwork (indicated by green box in (\textbf{a})) (ACC = 100\%, AUC = 1).
        (\textbf{d}) 
        The performance of PDIF-based reconstruction using the
        jittered spike trains (ACC = 99.99\%, AUC = 1).
        In (\textbf{b})-(\textbf{d}), $10^{7}\,\mathrm{ms}$ (around 2.7 hours) spike-train recordings
        (discrete time bin $\Delta t=0.5\,\mathrm{ms}$) are used for analysis, and 
        PDIF parameters are chosen as $k=l=1$, $\tau=3$ ms.
        (\textbf{e}) Schematic of the cerebellar circuit. Five major cell types and their principal projections 
        are illustrated for clarity.
        The full network comprises eight cell types (see Methods for details).
        (\textbf{f})-(\textbf{h}) PDIF reconstruction performance for inter-population projections: (\textbf{f}) PC–DCN (ACC = 95.49\%, AUC = 0.99), (\textbf{g}) DCN–IO (ACC = 98.96\%, AUC = 0.99), and (\textbf{h}) IO–DCN (ACC = 100.00\%, AUC = 1).
        Analysis is based on $10^{6}\,\mathrm{ms}$ (around 16.7 minutes) spike-train recordings
        with a bin size of $\Delta t=0.1\,\mathrm{ms}$).
        PDIF parameters are set to $k=1$ and $l=5$,
        and $\tau$ is listed in Tab.~\ref{tab:projection-params},  
        following the reconstruction pipeline shown in Fig.~\ref{fig:recon-pipeline}.
        In (\textbf{b})-(\textbf{d}) and (\textbf{f})-(\textbf{h}), orange and dark green histograms represent the distinct distributions
        of PDIF values for connected and unconnected pairs, respectively. The pink vertical line denotes the optimal reconstruction threshold determined via GMM (see Methods).
    }
    \label{fig:HH100}
\end{figure*}

Previously, we have demonstrated the underlying mechanism that enables successful PDIF reconstruction in small-scale networks. This theoretical framework naturally extends to larger network systems.
To further validate its applicability, we next evaluate the performance of PDIF on large-scale networks, highlighting its potential for real-world applications.
We simulate a 100-neuron HH network, randomly connected with 25\% connection density. In the simulation, we record network activity for $10^7$ ms (around $2.7$ hours), with the raster plot shown in Fig. \ref{fig:HH100}(a).
Following the reconstruction pipeline in Fig. \ref{fig:recon-pipeline}, we compute PDIF values of all neuronal pairs in the network, and the histogram (in log-scale) of PDIF values are shown in Fig. \ref{fig:HH100}(b). 
Here, we color the histogram of connected and unconnected pairs with orange and dark green, respectively, based on the ground truth connectivity (i.e., the adjacency matrix).
The PDIF values of these two groups are clearly separated, demonstrating PDIF's effectiveness
as a metric for discriminating connected neuronal pairs.
In practice, as the final step in the pipeline, we fit a double-component Gaussian mixture model (GMM)
to PDIF values, and then get the optimal threshold for reconstruction, as shown by the pink line in Fig. \ref{fig:HH100}(b).  
As a result, the structural connectivity can be fully reconstructed with $100\%$ accuracy.
As emphasized previously, PDIF implements pairwise reconstruction, inherently enabling the
reconstruction of subnetworks by utilizing partial observations of the overall network's activity dynamics.
Here, we randomly sample a 30-neuron subnetwork from the same 100-neuron network, indicated by the green box in Fig. \ref{fig:HH100}(a), 
and PDIF again achieves 100\% reconstruction accuracy, as shown in Fig. \ref{fig:HH100}(c).
We further evaluate the robustness of our PDIF reconstruction framework 
in the presence of measurement noise.
To introduce temporal fluctuations in spike timing,
we add uniformly distributed spike jitter ranging from $-2$ ms to $2$ ms into spike times,
as illustrated by the red bars in Fig. \ref{fig:HH100}(a). 
Such temporal noise partially corrupts the driving-driven relations between neuronal pairs.
Yet PDIF remains highly effective, achieving a reconstruction accuracy of 99.99\%, as shown in Fig. \ref{fig:HH100}(d). 

Furthermore, PDIF remains effective and robust when applied to general neuronal networks
characterized by biologically detailed neuronal models, heterogeneous connectivity, and quasi-synchronized network dynamics.
We first evaluate its performance in a cerebellar circuit network, which is built 
with multi-compartment neuronal model, and features both feedforward and feedback pathways of the circuit. The network follows
the established cerebellar architecture and
comprises eight distinct cell types, including Purkinje cells (PCs)~\cite{masoli2024human}, deep cerebellar nuclei (DCN) neurons~\cite{sudhakar2015cerebellar}, inferior olive (IO) neurons~\cite{zhang2019role}, etc.,
as illustrated in Fig.~\ref{fig:HH100}(e).
Detailed population sizes and cross-population projection parameters are
provided in Tab.~\ref{tab:cerebellar-pop} and~\ref{tab:projection-params}, respectively.
Overall, PDIF successfully reconstructs connections across 
different cell populations with an accuracy of 96.06\% $\pm$ 1.37\% and
an AUC of 0.98 $\pm$ 0.01.
To illustrate the reconstruction performance explicitly,
three representative projection pairs are
shown in Fig.~\ref{fig:HH100}(f)-(h), including PC-DCN, DCN-IO, and IO-DCN.
The PDIF distribution of connected and unconnected pairs forms distinct peaks, enabling reliable connectivity recovery.
Even with minor overlaps between distributions, PDIF maintains strong performance, with reconstruction accuracies of
95.49\%, 98.96\%, and 100.00\% and corresponding AUCs of 0.99, 0.99, and 1.00, respectively.
Reconstruction performances for the remaining projection pairs are detailed in Tab.~\ref{tab:cerebellar-perf} and Fig. \ref{fig:cerebellar_full_TE}.
We also test PDIF in networks with heterogeneous coupling strengths. 
Motivated by experimental observations that cortical synaptic weights follow a log-normal distributed coupling strength \cite{buzsaki2014logdynamic},
we simulate an HH network with log-normal distributed connection weights. PDIF achieves 99.26\% accuracy in distinguishing directly connected from unconnected pairs (Fig. \ref{fig:log-normal}(a)), and preserves the quadratic relation between PDIF values and coupling strengths (Fig. \ref{fig:log-normal}(b)).
Finally, PDIF remains effective under partial or moderate network synchronization induced by strong recurrent interactions 
or coherent external stimuli, where conventional causal inference methods often fail.
We simulate a 100-neuron HH network identical to that in Fig. \ref{fig:HH100}(b), except with coupling
strength $S = 0.028\,\mathrm{mS}\cdot\mathrm{cm}^{-2}$. 
The network exhibits quasi-synchronized firing (Fig.~\ref{fig:downsample}(a), top). 
To mitigate strong synchronous events, we apply a downsampling preprocessing step to remove strongly synchronized firing events
(red bars in Fig. \ref{fig:downsample}(a)).
PDIF reconstruction accuracy reaches 99.53\% on the downsampled data (Fig. \ref{fig:downsample}a, bottom),
as shown in Fig. \ref{fig:downsample}(b). A second example, where synchronization is driven by coherent Poisson inputs,
yields 99.93\% accuracy (Fig.~\ref{fig:downsample}(c)).

\subsection{Application of PDIF to general nonlinear networks}

As a model-free, information-theoretic measure, PDIF is broadly applicable to various types of nonlinear networks. To demonstrate its generality and robustness, we apply PDIF to four representative 100-node networks: the Lorenz network, logistic network, Rössler network, and random recurrent neural network (RNN). Each network has random connectivity with a density of 25\% (see Methods for details about network dynamics).
Notably, although these network models do not inherently generate spikes, their continuous-valued time series can be converted into surrogate spike trains by applying an appropriate threshold.

\begin{figure*}[htp]
    \centering
	\includegraphics[width=\textwidth]{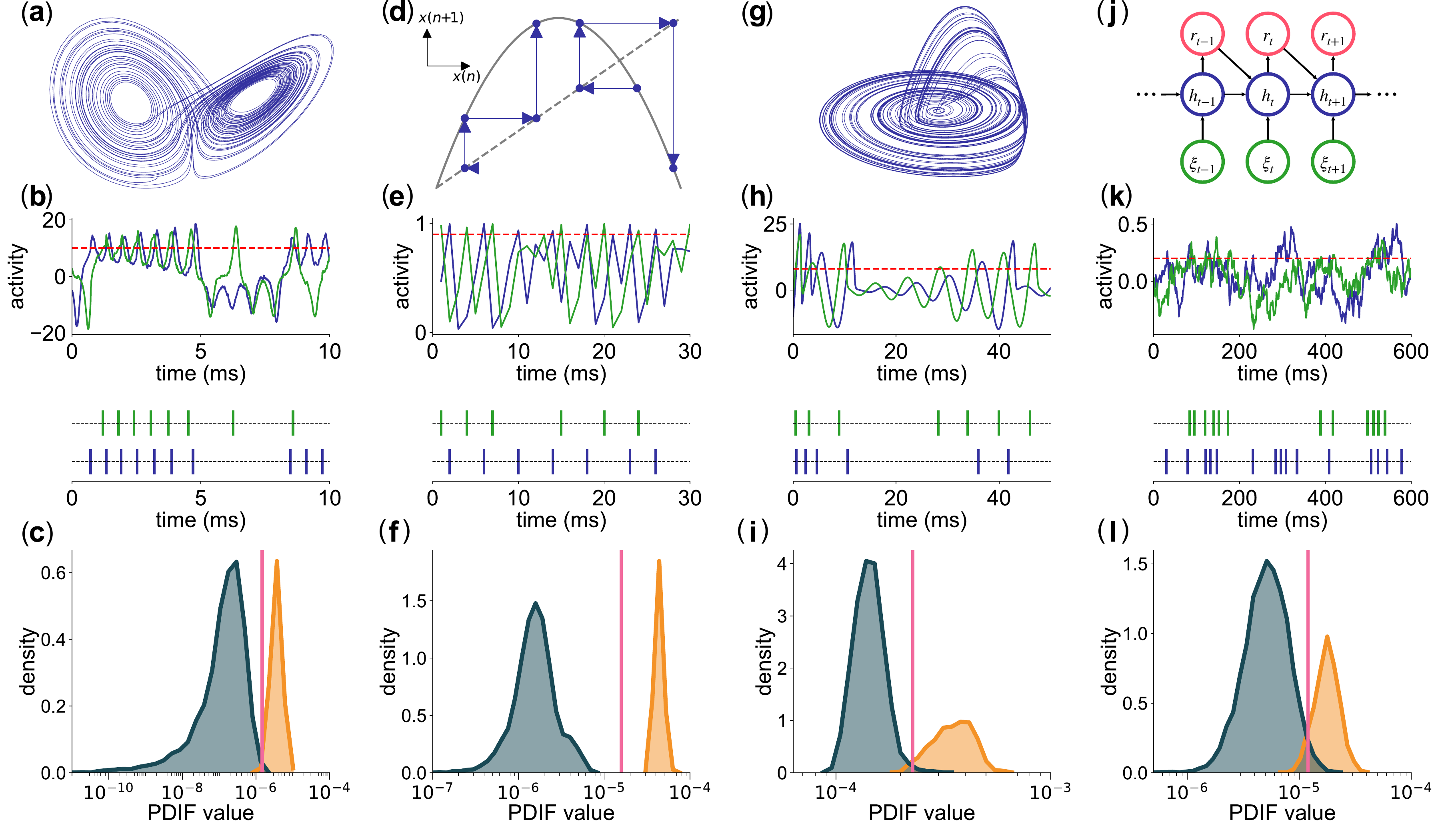}
	\caption{\textbf{PDIF reconstruct the connectivity of general nonlinear networks.}
    	(\textbf{a}) Demonstration of the node dynamics in Lorenz networks.
    	(\textbf{b})
        (Top) Continuous-valued activity time series of two example
        nodes from a 100-node Lorenz network.
        The red dashed line denotes the threshold value used for binarization 
        in the PDIF pipeline. 
        (Bottom) The corresponding surrogate spike trains for the same nodes.
        (\textbf{c}) The PDIF-based reconstruction performance for
        the same Lorenz network.
        The PDIF distributions for connected and unconnected pairs are colored by
        orange and dark green, respectively, with pink line indicating the optimal 
        reconstruction threshold determined by GMM (see Methods).
        (\textbf{d})-(\textbf{l}) Demonstration of binarization and reconstruction performance for ((\textbf{d})-(\textbf{f})) logistic networks,
        ((\textbf{g})-(\textbf{i})) Rössler networks, and ((\textbf{j})-(\textbf{l})) RNNs, respectively, organized similarly as (\textbf{a})-(\textbf{c}).
        PDIF parameters are chosen as
        $k=1$, $l=1$, $\tau=0$ ms for Lorenz networks,
        $k=2$, $l=1$, $\tau=0$ for logistic networks,
        $k=5$, $l=5$, $\tau=0$ ms for Rössler networks, and
        $k=5$, $l=1$, $\tau=5$ ms for RNNs.}
		\label{fig:general_nets}
\end{figure*}

In the Lorenz network, each node follows the Lorenz63 system \cite{lorenz2017deterministic}, which is governed by  
\begin{eqnarray*}
\dot{x}_i && =\sigma(y_{i}-x_{i})+S\sum_{j}A_{ij}\sum_{k}\delta(t-T_{jk}),\\
\dot{y}_i && =\rho x_i-y_i-x_iz_i,\\
\dot{z}_i && =-\beta z_i+x_iy_i,
\end{eqnarray*}
where $\sigma$, $\beta$ and $\rho$ are Lorenz parameters, and $S$ denotes the coupling strength between nodes (see details in Methods). $\mathbf{A}=(A_{ij})$ is its adjacency matrix, and $\delta (\cdot)$ is the Dirac delta function. 
When $x_j$ exceeds an interaction threshold $x_\theta$ from below at time $T_{jk}$,
the $j$-th node immediately sends its $k$-th pulse to all of its post nodes ($\{i|A_{ij}=1\}$).
In Fig. \ref{fig:general_nets}(a), we show the 3D state space trajectory for one of the Lorenz nodes in coupled network.

We record the time series of the $x$-variable from each node and apply a binarization threshold of $x^\mathrm{th} = x_\theta$ --matching the interaction threshold -- (red dashed line in Fig. \ref{fig:general_nets}(b), top) to generate surrogate spike trains (Fig. \ref{fig:general_nets}(b), bottom).
The resulting binary time series, discretized at a time step $\Delta t = 0.02\,\mathrm{ms}$, are then passed as input to the PDIF pipeline with parameters $k = l = 1$ and $\tau = 0\,\mathrm{ms}$.
As shown in Fig. \ref{fig:general_nets}(c), PDIF values for connected (orange) and unconnected (dark green) node pairs show clear separation, enabling reconstruction with 99.81\% accuracy.
The receiver operating characteristic (ROC) curve yields an area under the curve (AUC) of 0.99, indicating high reconstruction performance.

For the logistic network, each node evolves according to the discrete-time logistic map \cite{may1976simple,dickten2014identifying}, as shown in Fig. \ref{fig:general_nets}(d). Its dynamics can be depicted as a discrete dynamical system, governed by 
\begin{equation*}
x_{i}(n+1)=rx_{i}(n)[1-x_{i}(n)]+S\sum_{j}A_{ij}\sum_{k}\tilde{\delta}_{n+1,T_{jk}},
\end{equation*}
where $r$ is the parameter of logistic map and $S$ is the coupling strength (see details in Methods). 
Similarly, $\mathbf{A}=(A_{ij})$ is its adjacency matrix, and $\tilde{\delta}$ is the Kronecker delta function.
$T_{jk}$ is the interaction time stamp satisfying $x_{j}(T_{jk}-1)<x_{\theta}$ and $x_{j}(T_{jk})\geq x_{\theta}$, where $x_{\theta}$ is the interaction threshold. 
Naturally, we take the sequence of interaction time stamps $\{T_{jk}\}$ for each node
as their surrogate spike trains
({i.e.}, binarization threshold $x^\mathrm{th}=x_\theta$ and discrete time step $\Delta t=1$)
for PDIF analysis. With PDIF parameters $k = 2$, $l = 1$, and $\tau = 0$, the method yields perfect reconstruction (Fig. \ref{fig:general_nets}(f)), achieving 100\% accuracy and an AUC of 1.

In the Rössler network, node dynamics follow the diffusive-coupled Rössler system \cite{rossler1976equation}, whose dynamics are defined by
\begin{eqnarray*}
\dot{x_{i}} &&= - y_i - z_i + S\sum_{j \neq i}^{N} A_{i j}  \cdot\left(x_{j}-x_{i}\right) \\ 
\dot{y_{i}} &&= x_{i}+ay_{i} \\ 
\dot{z_{i}} &&=-b + z_{i}(x_{i}-c) 
\end{eqnarray*}
where $a$, $b$ and $c$ are parameters of the Rössler system,
and $S$ is the coupling strength (see details in Methods).
Similarly, $\mathbf{A}=(A_{ij})$ is its adjacency matrix.
The 3D state space trajectory for one of the Rössler node in coupled network is shown in Fig. \ref{fig:general_nets}(g).
We use the time series of the $x_i(t)$ component for reconstruction, applying a threshold $x^\mathrm{th} = 8$ (Fig. \ref{fig:general_nets}(h)) and a binarization step of $\Delta t = 3\,\mathrm{ms}$. Using PDIF parameters $k = 5$, $l = 5$ and $\tau = 0$ ms, the method achieves 98.61\% reconstruction accuracy, with an AUC of 0.99 (Fig. \ref{fig:general_nets}(i)).

Lastly, we test PDIF's performance in the random recurrent neural network (RNN) \cite{sompolinsky1988chaos}. The dynamics of the $i$-th node of a noise-driven random
recurrent neural network is described by
\begin{equation*}
\tau_m \dot{x}_i = -x_i + S\sum_{j}A_{ij} \tanh(x_{j}) + \xi_i,
\end{equation*}
where $\tau_m$ is the time constant, and $S$ is the coupling strength. 
$\mathbf{A}=(A_{ij})$ is its adjacency matrix,
and $\xi_i$ is the independent Gaussian white noise for each node (see details in Methods).
We simulate the RNN dynamics using forward Euler scheme, as shown in Fig. \ref{fig:general_nets}(j).
After simulating the network, we apply a threshold of $x^\mathrm{th} = 0.2$ to the node states (Fig. \ref{fig:general_nets}(k)) and binarize with $\Delta t = 5\,\mathrm{ms}$. Using parameters $k = 4$, $l = 1$, and $\tau = 5\,\mathrm{ms}$, PDIF achieves 93.55\% reconstruction accuracy (Fig. \ref{fig:general_nets}(l)), with an AUC of 0.98.

% \clearpage

\subsection{Comparison with other network reconstruction methods}

To further demonstrate the advantages of our method over other existing approaches, we next compare PDIF reconstruction with various popular methods for nonlinear network reconstruction, including
symbolic transfer entropy (STE) \cite{staniek2008symbolic},
GLM for cross correlation (GLMCC) \cite{kobayashi2019reconstructing}, dynamical differential covariance (DDC) \cite{chen2022dynamical}, convergent cross mapping (CCM) \cite{sugihara2012detecting} and its related variations as frequency-domain CCM (FDCCM) and symbolic CCM (SCCM) \cite{avvaru2023effective, ge2023symbolic}.
For fair comparison, all other methods are implemented in a pairwise manner, except for DDC which relies on estimation of the inverse of network correlation matrix.
Notably, although conditional versions exist for some methods (in a format analogous to conditional TE), they also suffer from the curse of dimensionality and are thus not applicable in practice. We construct 8 different types of nonlinear networks and each includes 10 nodes with 25\% connection density, including excitatory HH networks (HHEE),
excitatory-inhibitory mixed HH networks (HHEI),
excitatory continuously coupled HH networks (HHconEE),
excitatory-inhibitory mixed, continuously-coupled HH networks (HHconEI),
Lorenz networks,
logistic networks,
Rössler networks,
and RNNs.

\begin{figure*}[htp]
    \centering
	\includegraphics[width=0.9\textwidth]{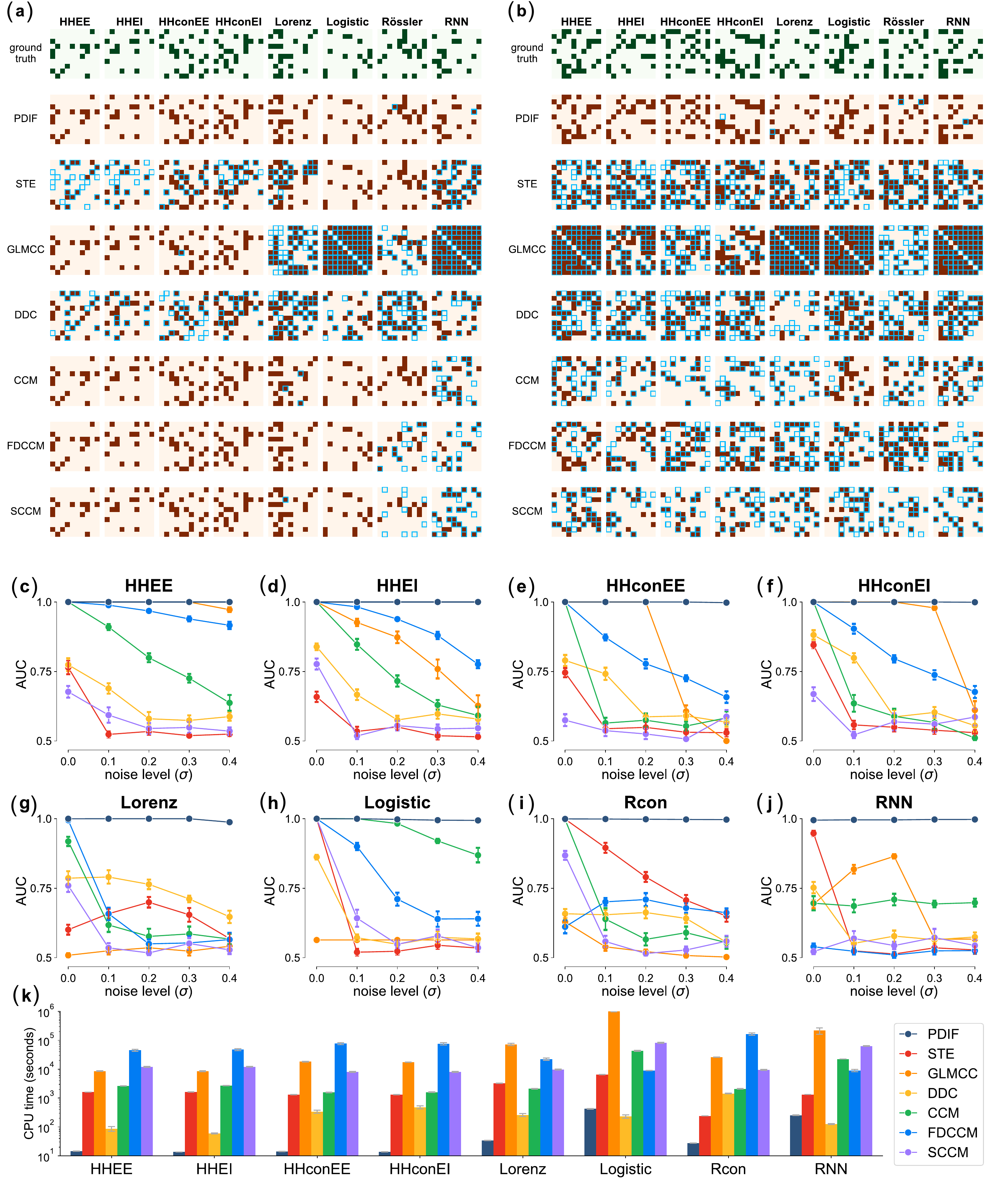}
	\caption{\textbf{Reconstruction performances of multiple networks using different methods.}
    (\textbf{a}) Reconstructed adjacency matrices for 10-node networks.
    The first row shows ground truth adjacency 
    matrices for 8 benchmark networks, followed by reconstructed adjacency matrices
    (one method per line). Blue squares indicate
    either false positives or negatives. Reconstruction thresholds are
    determined by fitting GMMs to metrics (see Methods).
    (\textbf{b}) Reconstructed adjacency matrices for 10-node subnetworks with measurement
    noise. Layout and color scheme match those in (\textbf{a}).
    (\textbf{c})-(\textbf{j}) 
    Reconstruction performances (AUCs) changes with noise levels in the
    nodes' activity measurement of continuous-valued time series for
    the 8 benchmark network systems, respectively.
    For each network system, the measurement noise is calibrated in units of 
    standard deviation of its own noise-free 
    continuous-valued time series.
    The reconstructions in (\textbf{b}) correspond to noise level $\sigma=0.3$.
    Each data point is the result average among 10 subnetworks randomly sampled from a large network, with the error bar indicates the standard error of mean.
    (\textbf{k}) The average CPU time costs for each reconstruction methods to reconstruct different networks. The error bars indicate the standard error of mean among all reconstructions among different noise levels and trials.
    PDIF parameters follow the standard reconstruction pipelines (Fig. \ref{fig:recon-pipeline}).}
    \label{fig5}
\end{figure*}

We evaluate the reconstruction performances of all methods using reconstruction accuracy and AUC values.
As shown in Fig. \ref{fig5}(a), PDIF reconstruction consistently outperforms other methods in all 8 
benchmarks, with AUC values being or approaching unity. To provide an intuitive visualization of reconstruction performance,
incorrectly inferred connections are marked by blue squares in Fig. \ref{fig5}(a) for all methods.
DDC performs worst as it requires an accurate estimation of the derivative of voltage time series, and is more sensitive to the detection for inhibitory interactions. Among other methods, CCM and its variations achieve comparable reconstruction performances as PDIF in many benchmarks but not the weakly coupled RNN system, where their noise sensitivity degrades accuracy. 

The advantage of PDIF becomes more substantial when we evaluate the performance of these methods by reconstructing 10-node subnetworks, sampled from 100-node large networks, using noisy measurements of node activities. As shown in Fig. \ref{fig5}(b), PDIF demonstrates the feasibility of subnetwork reconstruction and again consistently and dominantly outperforms other methods. Unlike that of other methods, which degrades with increasing measurement noise, PDIF reconstruction possesses the robustness of performance against noisy measurements  (Fig. \ref{fig5}(c)-(j)). Furthermore, the computational cost of PDIF is primarily determined by the data loading and estimation of the joint probability distribution,
which makes PDIF more efficient than all other reconstruction methods (Fig. \ref{fig5}(k) and Fig. \ref{fig:wall-time}).

% \clearpage

\subsection{Application of PDIF on real spike-train data}

To investigate the applicability of PDIF to real data analysis, we now apply PDIF to spike trains recorded from multiple regions of mice cortex in order to reconstruct the underlying neuronal connectivity. Here we analyze the \textit{Visual Coding Neuropixels} dataset and \textit{Visual Behavior - Neuropixels} dataset
that are publicly available from \textit{Allen Brain Observatory} \cite{siegle2021survey,
allendata}.
The former surveys spike trains mainly in visual cortices during passive viewing of visual stimuli,
while the latter records the neuronal activity in mice performing the visual change detection tasks.
We analyze spike-train recordings from 135 sessions (32 visual coding sessions and 103 visual behavior sessions).
For each session, neurons with firing rates exceeding $0.05$ Hz and signal-to-noise ratios above $4$ are
included. Spike trains are extracted under different stimulus or behavior conditions for separate reconstructions.
In visual coding sessions, spike trains during passive viewing of drifting gratings, static gratings, natural scenes, and natural movies (each $>25$ minutes) are analyzed. In visual behavior sessions, active and passive behavior periods (each $>1$ hour) are examined separately.

\begin{figure*}[htp]
    \centering
    \includegraphics[width=0.9\textwidth]{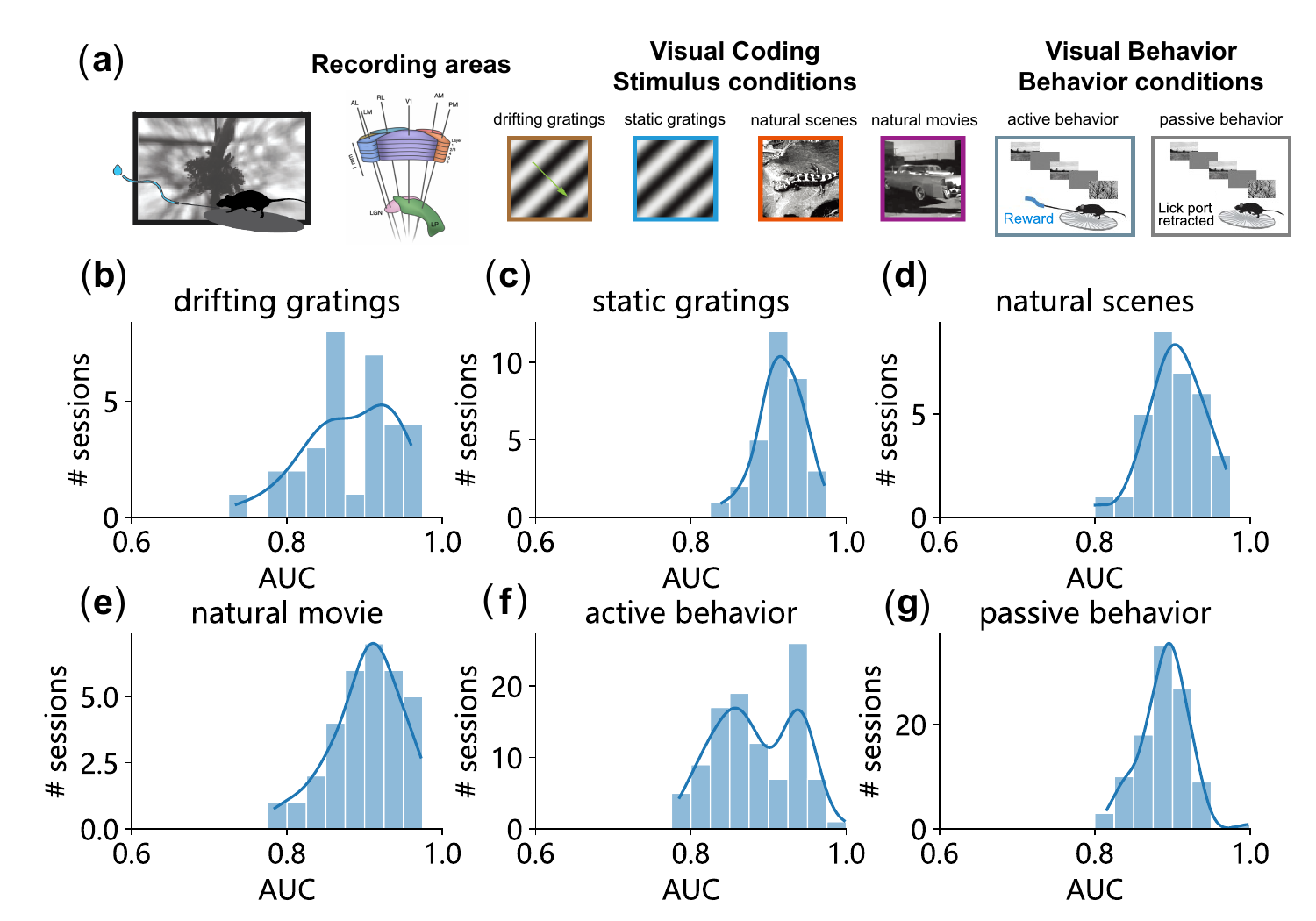}
    \caption{
    \textbf{Performance of PDIF reconstruction using spike trains from mice data.}
    \textbf{a}. Schematics of the behaving mouse, recording cortical areas, and six different recording conditions 
    (four stimulus condition in visual coding dataset and two types of behaviour states in visual behavior dataset).
    (\textbf{b})-(\textbf{e}) Histograms of AUC values for reconstruction across sessions in the visual coding dataset 
    for (\textbf{b}) drifting gratings, (\textbf{c}) static gratings, (\textbf{d}) natural scenes, and (\textbf{e})
    natural movie, respectively.
    (\textbf{f})-(\textbf{g}) Histograms of AUC values for reconstructions across sessions in the visual behavior dataset
    under (\textbf{f}) active and (\textbf{g}) passive conditions.
    Blue lines indicate kernel density estimates of the histograms.
    }
    \label{fig6}
\end{figure*}

Due to experimental limitations, ground-truth structural connectivity is unavailable.
Thus, we quantify reconstruction performance using AUCs from double-component GMM fits to log-scaled
PDIF values (see Methods). As shown in Fig. \ref{fig6}, the peak of AUC value distributions are around 0.9 for most sessions. Assuming structural connectivity of the neuronal network in the mice cortex remains stable within a session, we expect consistent reconstructed connectivity across conditions. Cross-condition consistency of reconstructed network exceeds 74\% for the visual coding dataset and 69\% for the visual behavior dataset, further supporting PDIF’s validity for real spike-train analysis.

% \clearpage

\section{Discussion}\label{sec:discussion}

In this work, we have introduced a general, model-free, and scalable network reconstruction pipeline
that quantifies pairwise information flow with explicit time delays.
The proposed framework effectively circumvents the curse of dimensionality inherent in conventional information-theoretic approaches, which arises from the requirement of estimating high-dimensional probability distributions in the course of reconstructing network structural connectivity.
We also revealed a quadratic relationship between PDIF value and the coupling strength between nodes, providing a theoretical foundation for PDIF-based network reconstruction. We further demonstrated the orders of magnitude difference between PDIF values for connected and unconnected pairs, enabling accurate structural connectivity reconstruction. Our results showed that PDIF successfully reconstructs HH neuronal networks, as well as other nonlinear networks, including Lorenz networks, logistic networks, Rössler networks and random recurrent neural networks, validating its underlying mechanism. Through comparisons with other methods across various networks, we demonstrated that PDIF consistently outperforms other methods, exhibiting its broad applicability to subnetwork reconstruction and robustness to noisy activity measurements. Finally, we applied PDIF to real spike-train data recorded from electrophysiological experiments. High AUC values across all sessions, along with strong cross-condition consistency, further validate PDIF's applicability to real experimental data.

It would be ideal to measure network structural connectivity directly. In neuroscience, the golden standard for assessing cellular-level structural connectivity involves techniques such as electron microscopic reconstruction of inter-neuronal synaptic structures \cite{themicronsconsortium2025functional} or virus tracing 
\cite{xu2021highthroughput}. However, these methods are either time- and resource-intensive or limited in their ability to reconstruct large numbers of neurons.
In contrast, recent advances in electrophysiology and neural imaging, such as Neuropixel recordings \cite{steinmetz2021neuropixels} and large-scale calcium imaging \cite{grienberger2022twophoton}, 
enable simultaneous cellular-level activity recordings from large neuron populations in behaving animals. These advancements open the possibility of inferring network structures using reconstruction methods, underscoring the importance of this work.

Compared to other types of neuronal recordings, spike-train data offers a high signal-to-noise ratio. Motivated by this, PDIF was initially developed for spike-train-based reconstructions and later extended to general nonlinear systems. Unlike other methods focused on binary time series, PDIF fully leverages the advantageous properties of binary data in several ways. 
First, the binary state and short memory favor small order parameters, which reduces the dimensionality of PDIF estimation. Second, the direct interactions between binary time series ensure robustly large PDIF values against the effects of indirect interactions, enabling feasible pairwise reconstruction and enhancing its suitability for real-world applications. Third, unlike DDC and CCM, which are applied to continuous-valued time series and require smoothness in the recorded data, binarization acts as a filter for noisy subthreshold fluctuations, making PDIF robust to noisy measurements, as illustrated in Fig. \ref{fig5}(c).  
When applied to other general nonlinear networks, a similar binarization process is employed with an appropriately chosen threshold for continuous-valued time series. As shown in Fig. \ref{fig5}, PDIF is broadly extendable to various physical networks, consistently achieving robust and reliable reconstruction performance.

While PDIF necessitates relatively long spike recordings to achieve effective reconstruction — a reasonable trade-off for the substantial noise-filtering benefits conferred by binarization — the data requirements remain within experimentally accessible bounds. Specifically, accurate reconstruction of a 100-neuron Hodgkin–Huxley network is achievable with approximately $10^6$ ms of spike recordings (Fig. \ref{fig:HH100_TE_vs_length}), a duration that is feasible in experimental settings; the extended $10^7$ ms recordings employed in Fig. \ref{fig:HH100}(a) reflect a conservative choice prioritizing reconstruction fidelity. Nevertheless, reconstruction performance degrades with reduced data length (Fig. \ref{fig:HH100_TE_vs_length}), identifying the development of data-efficient inference strategies as a productive direction for future methodological refinement. A further consideration arises in networks characterized by high connection density or strong coupling, wherein the cumulative contribution of indirect causal interactions may challenge reconstruction accuracy: although PDIF values associated with indirect pathways are orders of magnitude smaller than those of direct interactions, they can aggregate across an increasing number of indirect paths (Fig. \ref{fig:HH100_indirect}). Compounding this, denser networks tend to exhibit greater correlated and synchronized activity, which itself poses well-documented challenges for connectivity reconstruction \cite{das2020systematic}. Addressing these regimes represents a natural and well-motivated future extension of the PDIF framework.

\section{Methods}\label{sec:methods}

\subsection{PDIF network reconstruction pipeline}

\noindent\textbf{Step 1}: Convert continuous-valued time series, $x(t)$, to binary time series, $S^x(n)$.  
For a continuous-valued time series $x(t)$, if $x(\hat{t}) = x^{\mathrm{th}}$ and $\dot{x}(\hat{t}) > 0$, we define $T_k = \hat{t}$ as the spike time of the $k$-th pulse event, where $x^{\mathrm{th}}$ is the binarization threshold. Using a discrete time step $\Delta t$, we construct a binary time series $S^x(n)$, where $S^x(n) = 1$ if a pulse event occurs within $((n-1)\Delta t, n\Delta t]$, otherwise $S^x(n) = 0$. $\Delta t$ is chosen such that at most one pulse appears within each time step. %Inspired by action potential generation, we define pulse-output events for general continuous-valued time series using a threshold.  

\noindent\textbf{Step 2}: Determine the order parameter $k$ (for the target node).  
For a $k$-th order Markov process, $k$ removes the contribution of information transfer from the target’s own history. In practice, $k$ is chosen as the time lag
$\hat{k}$ where the absolute value of the autocorrelation function (ACF) of $S^x(n)$ drops below 0.1, i.e., 
\begin{equation*}
\hat{k}=\arg\min\limits_k\{|\mathrm{ACF}|_{S^x(n)}(k)<0.1\}.
\end{equation*}

\noindent\textbf{Step 3}: Determine the optimal time delay $\tau$.  
Scan and choose a positive $\tau$ as $\hat\tau$ which maximize 
the time-delayed mutual information (TDMI) from $X$ to $Y$, {i.e.}, $\hat{\tau}=\arg\max_\tau MI_{X\to Y}(\tau)$, where 
\begin{eqnarray*}
MI_{X\to Y}(\tau)=\sum\limits_{y_{n+1},x_{n-\tau}}
p\left(y_{n+1},x_{n-\tau}\right)\\
\cdot\log\frac{p\left(y_{n+1}\right)}{p\left(y_{n+1}\right)p\left(x_{n-\tau}\right)}.
\end{eqnarray*}
Notably, TDMI can be regarded as a simplified PDIF measure that 
excludes the target's history and uses a single source-history term
($x_{n-\tau}^{(l)}\to x_{n-\tau}$). This delay $\hat{\tau}$ characterizes the dominant interaction
timescale.

\noindent\textbf{Step 4}: Determine the order parameter $l$ (for the source node).  
Given the optimal $\tau$, $l$ is chosen by scanning values from 1 to 10. In practice, $l = 1$ is sufficient for most systems, with larger values used only for dynamics with slow, persistent interactions (in this work, $l = 1$ for all networks except for cerebellar networks, Rössler systems and real spike trains, for which $l = 5$). 

\noindent\textbf{Step 5}: Determine the classification threshold.  
After computing PDIF values for all pairs in a network using these parameters chosen in Steps 1–4, we fit a double-component Gaussian mixture model to the resulting values and select the classification threshold at the point where the probabilities of the two mixtures are equal, treating the mixtures as proxies for connected and unconnected pairs.

\subsection{Reconstruction performance criteria for simulation data}

For each reconstruction method, we infer the network adjacency matrix by applying an optimal threshold to its estimated causality values, obtained by fitting a Gaussian mixture model to the distribution of values across all node pairs. Reconstruction accuracy is defined as the agreement between the inferred adjacency matrix and the ground truth connectivity. To assess sensitivity and specificity, we compute receiver operating characteristic (ROC) curves and use the area under the curve (AUC) to quantify how well each method separates connected from unconnected pairs. An AUC near 1 indicates clear separability and thus near-perfect reconstruction, whereas an AUC close to 0.5 reflects substantial overlap and performance similar to random guessing. Some methods, including GLMCC and DDC, produce both positive and negative metrics, with negative values often interpreted as inhibitory interactions. Because our evaluation concerns only the presence or absence of a connection, we compute all performance metrics using the absolute values of these measures.

\subsection{Reconstruction performance criteria for real spike-train data}

When working with real spike-train data, the true structural connectivity is unknown, preventing direct evaluation of reconstruction performance. Prior experiments indicate that coupling strengths in mouse and monkey cortex follow a log-normal distribution \cite{buzsaki2014logdynamic}. Guided by this observation, we fit a Gaussian mixture model to the log-transformed PDIF values, treating the two resulting components as proxies for the distributions of connected and unconnected pairs. Using these fitted distributions, we compute ROC curves and their AUC values to quantify how well PDIF separates the two classes, as shown in Fig.~\ref{fig6}.

\subsection{Network models}
\subsubsection{Hodgkin-Huxley neuronal network}

The dynamics of the $i$-th Hodgkin-Huxley (HH) neuron's membrane
potential $V_i(t)$ is governed by 
\begin{eqnarray}\label{eq:HH}
C\dot{V}_i =&& -(V_{i}-V_\mathrm{Na})G_\mathrm{Na}m_{i}^{3}h_{i}-(V_{i}-V_\mathrm{K})G_\mathrm{K}n_{i}^{4} \nonumber\\
&&-(V_{i}-V_\mathrm{L})G_{L}-(V_{i}-V_\mathrm{F})G_{i}^{\text{F}} \\
&&- (V_{i}-V_\mathrm{E}) G_{i}^{\text{E}}
- (V_{i}-V_\mathrm{I}) G_{i}^{\text{I}},\nonumber
\end{eqnarray}
where $C$ is the membrane capacitance. $V_\mathrm{Na},V_\mathrm{K}$, and $V_\mathrm{L}$
are the reversal potentials for the sodium current, potassium current, and leak current, respectively.
$G_\mathrm{Na}$, $G_\mathrm{K}$, and $G_\mathrm{L}$
are the corresponding maximum conductance of those ionic trans-membrane currents.
To capture the nonlinear gating effect of sodium and potassium ion channel, HH neuron has three gating variables, $m_{i}$, $h_{i}$, and $n_{i}$, whose dynamics are defined by 
\begin{equation}
    \dot{z}_i =(1-z_{i})\alpha_{z}(V_{i})-z_{i}\beta_{z}(V_{i}), \,\,\, \text{ for }z=m,h,n.
\end{equation}
The forms of $\alpha_z$ and $\beta_z$ are defined by
\begin{eqnarray}
\alpha_{m}(V)&&=(0.1V+4)/[1-\exp(-0.1V-4)], \nonumber\\
\beta_{m}(V)&&=4\,\exp[-(V+65)/18],\nonumber\\
\alpha_{h}(V)&&=0.07\,\exp[-(V+65)/20], \nonumber\\
\beta_{h}(V)&&=1/[1+\exp(-3.5-0.1V)],\\
\alpha_{n}(V)&&=(0.01V+0.55)/[1-\exp(-0.1V-5.5)], \nonumber\\
\beta_{n}(V)&&=0.125\,\exp[-(V+65)/80], \nonumber
\end{eqnarray}
The input synaptic current received by the HH neuron can be split into 
three parts, including feedforward input from the homogeneous Poisson
processes, excitatory and inhibitory recurrent interactions from other neurons in the network.
$V_\mathrm{F}$, $V_\mathrm{E}$, and $V_\mathrm{I}$ are the corresponding reversal potential of those three types of current. $G_i^\mathrm{F}$, $G_i^\mathrm{E}$, and $G_i^\mathrm{I}$
are their corresponding conductances, which are defined by
\begin{eqnarray}
G_{i}^{\text{F}}(t)&&=f\sum_{k}H(t-T_{ik}^{\text{F}}), \nonumber\\
G_{i}^{\text{E}}(t)&&=\sum_{j}^{N_E}A_{ij}S^{Q_iE}\sum_{k}H(t-T_{jk}),\label{eq:dynG}\\
G_{i}^{\text{I}}(t)&&=\sum_{j}^{N_I}A_{ij}S^{Q_iI}\sum_{k}H(t-T_{jk}),\nonumber
\end{eqnarray}
where $Q_i\in\{E,I\}$ denotes the type of the $i$-th neuron.
Here, $f$ is the strength of the Poisson process with rate $\nu$, and 
$T_{ik}^{\text{F}}$ is the arrival time of the $k$-th Poisson spike for 
the $i$-th neuron. The temporal course of conductance change induced by
a single Poisson spike is modeled by a double exponential function
\begin{equation}
H(t)=\frac{\sigma_{r}\sigma_{d}}{\sigma_{d}-\sigma_{r}}[\text{exp}(-t/\sigma_{d})-\text{exp}(-t/\sigma_{r})]\Theta(t),
\label{eq:doubleexp}
\end{equation}
where $\sigma_{r}$ and $\sigma_{d}$ are  the rise and decay timescales, 
respectively, and $\Theta(t)$ is the Heaviside function.
The recurrent inputs are constrained by the structural connectivity
of the network, defined by adjacency matrix $\mathbf{A}=(A_{ij})$,
and the cell-type specific coupling strength $S^{Q_iE}$ and $S^{Q_iI}$
($Q_i$ is the type of the $i$-th neuron).
To be consistent with the notation, the coupling strength from $j$-th neuron to $i$-neuron defined in the main text 
is $S_{ij}=A_{ij}S^{Q_iE}$ for excitatory connections ($S_{ij}=A_{ij}S^{Q_iI}$ for inhibitory connections).
$T_{jk}$ is the arrival time of the $k$-th spike from the $j$-th neuron.
The dynamics of spike-induced conductance evolution is also
governed by Eq. (\ref{eq:doubleexp}). 
When the voltage $V_{i}$, described by Eqs. 
(\ref{eq:HH})-(\ref{eq:doubleexp}), exceeds the threshold $V^{\textrm{th}}$ at $T_{ik}$, i.e.,
\begin{equation*}
V_i(T_{ik})=V^\mathrm{th}\quad\mathrm{and}\quad \dot{V}_i(T_{ik}) > 0,
\end{equation*}
this event is denoted as the $k$-th spike of the $i$-th neuron. This triggers synaptic inputs to all its downstream neurons, as defined by Eqs. (\ref{eq:dynG})-(\ref{eq:doubleexp}).  

In this work, we choose parameters for simulating HH networks as listed below:
$C=1\,\mu\mathrm{F}\cdot\mathrm{cm}^{-2}$,
$V_\mathrm{Na}=50$ mV, $V_\mathrm{K}=-77$ mV, $V_\mathrm{L}=-54.387$ mV, $G_\mathrm{Na}=120\,\mathrm{mS\dots cm}^{-2}$,
$G_\mathrm{K}=36\,\mathrm{mS\cdot cm}^{-2}$,  $G_\mathrm{L}=0.3\,\mathrm{mS\dot cm}^{-2}$,
$V_\mathrm{F}=V_\mathrm{E}=0$ mV, $V_\mathrm{I}=-80\,\mathrm{mV}$, $V^{\mathrm{th}}=-50$ mV,
$\sigma_{r}=0.5$ ms, $\sigma_{d}=3.0$ ms, $\nu=0.15\,\mathrm{ms}^{-1}$, $f=0.08\,\mathrm{mS\cdot cm}^{-2}$, 
$S^{EE}=S^{IE}=0.02\,\mathrm{mS\cdot cm}^{-2}$, and 
$S^{EI}=S^{II}=0.08\,\mathrm{mS\cdot cm}^{-2}$.
Time step $\Delta t=0.5$ ms is used in the binarization step of PDIF reconstruction.
The typical data length for reconstructing a HH network is around 2.7 hours ($10^{7}\,\mathrm{ms}$) in Fig.~\ref{fig:HH100}.

\subsubsection{Cerebellar network}
We constructed a cerebellar network model based on the classical Hodgkin-Huxley (HH) framework, adopting established models from previous studies for granule cells (GrC)~\cite{masoli2020parameter}, Golgi cells (GoC)~\cite{masoli2020cerebellar}, basket cells (BC)~\cite{masoli2025cerebellar}, stellate cells (SC)~\cite{rizza2021stellate}, 
Purkinje cells (PC)~\cite{masoli2024human}, deep cerebellar nucleus (DCN) neurons~\cite{sudhakar2015cerebellar}, inferior olive (IO) neurons~\cite{zhang2019role}, and mossy fibers (MF) represented as Poisson spike generators, comprising eight cerebellar neuronal populations in total.
The population sizes and baseline firing rates of these neurons are summarized in Tab.~\ref{tab:cerebellar-pop}. 
The network connectivity and detailed synaptic parameters for all projections are provided in Tab.~\ref{tab:projection-params}. The membrane potential threshold used for binarization is set to $V^\mathrm{th}=0 \mathrm{mV}$.
In this work, a time series length of $17$ minutes ($10^6\,\mathrm{ms}$) is used to reconstruct the cerebellar network, with PDIF parameters set as $k=1$, $l=5$, and a binning time step $\Delta t=0.1$ ms
(see Figs.~\ref{fig:HH100} and \ref{fig:cerebellar_full_TE}). The optimal time delay $\tau$ for each projection, along with the corresponding reconstruction accuracy (ACC) and AUC, are reported in Tab.~\ref{tab:cerebellar-perf}.

\subsubsection{Continuously coupled Hodgkin-Huxley neuronal network}
The dynamics of the $i$-th continuously coupled HH neuron's
membrane potential follows the same set of equations as in classical HH network (Eqs.~(\ref{eq:HH})-(\ref{eq:doubleexp})),
with the conductance of recurrent synaptic inputs (in Eq. (\ref{eq:doubleexp})) replaced by the smooth function 
\cite{compte2003cellular,sun2010pseudolyapunov}
\begin{equation}
    \begin{aligned}
        G^E_i &=\sum_j^{N_E}A_{ij}S^{Q_iE}\frac{1}{1+\exp(-0.5V_{j}+10)},\\
        G^I_i &=\sum_j^{N_I}A_{ij}S^{Q_iI}\frac{1}{1+\exp(-0.5V_{j}+10)},\\
    \end{aligned}
    \label{eq:HHcon}
\end{equation}
where $Q_i\in\{E,I\}$ denotes the type of the $i$-th neuron.
Other parameters for continuously coupled HH networks are the same as those used in pulse-coupled
HH networks.
The threshold used for binarization is chosen as the same firing threshold $V^\mathrm{th}=-50\,\mathrm{mV}$ as that in HH networks.
In this work, a time series length $2.7$ hours ($10^7\,\mathrm{ms}$) is used
to reconstruct continuously coupled HH networks,
with PDIF parameters set as $k=1$, $l=1$, $\tau=3$ ms, and a binning time step $\Delta t=0.5$ ms,
following the reconstruction pipeline (see Fig.~\ref{fig5}).

\subsubsection{Lorenz network}
The dynamics of the $i$-th node in the pulse-coupled Lorenz networks is governed by \cite{lorenz2017deterministic, belykh2006synchronization, staniek2008symbolic, palus2007directionality} 
\begin{eqnarray}
\dot{x}_i && =\sigma(y_{i}-x_{i})+S\sum_{j}A_{ij}\sum_{k}\delta(t-T_{jk}),\nonumber\\
\dot{y}_i && =\rho x_i-y_i-x_iz_i,\\
\dot{z}_i && =-\beta z_i+x_iy_i,\nonumber
\label{eq:Lorenz}
\end{eqnarray}
where $\sigma=10$, $\beta=8/3$, $\rho=28$, $S=0.5$, and $\delta (\cdot)$ is the Dirac delta function. 
$\mathbf{A}=(A_{ij})$ is its adjacency matrix, with a connection density of 25\% throughout this work.
When $x_j$ exceeds the threshold $x_\theta=10$ from below at time $T_{jk}$, node $j$ immediately sends its $k$-th pulse to all of its post nodes ($\{i|A_{ij}=1\}$).
Due to this pulse-coupled property, we naturally choose the same $x^\mathrm{th}=x_\theta$ as the 
binarization threshold in PDIF reconstruction pipeline, with a binning time step $\Delta t=0.02$ ms.
In this work, time series of $x$-variable with length $17$ minutes 
($10^6\,\mathrm{ms}$) are used to reconstruct Lorenz networks.
PDIF parameters are chosen as $k=1$, $l=1$, and $\tau=0$ ms according to the reconstruction pipeline (see Figs~\ref{fig:general_nets} and \ref{fig5}).

\subsubsection{Logistic network}
The dynamics of the $i$-th node in logistic networks are governed by the discrete time logistic map \cite{may1976simple,dickten2014identifying}
\begin{equation}
x_{i}(n+1)=rx_{i}(n)[1-x_{i}(n)]+S\sum_{j}A_{ij}\sum_{k}\tilde{\delta}_{n+1,T_{jk}},
\end{equation}
where $r=4$, $S=0.005$, and $\tilde{\delta}$ is the Kronecker delta function.
$\mathbf{A}=(A_{ij})$ is its adjacency matrix, with a connection density of 25\% throughout this work.
We record $T_{ik}$ as the discrete time point when $x_{i}(T_{ik}-1)<x_\theta$ and $x_{i}(T_{ik})\geq x_{\theta}$, where $x_{\theta}=0.9$ is the interaction threshold.
To apply PDIF, we convert $x_i(n)$ to binary time series by setting the interaction threshold as the 
binarization threshold of $x_i(n)$, and choosing a binning time step $\Delta t = 1$. 
For reconstructions shown in this work, 
PDIF parameters are chosen as $k=2, l=1, \tau=0$, and time series of $x_i(n)$
with length $10^7$ time points are used for reconstruction (see Figs~\ref{fig:general_nets} and \ref{fig5}). 

\subsubsection{Rössler network}
The dynamics of the $i$-th node of a diffusive-coupled Rössler system is defined by \cite{rossler1976equation, staniek2008symbolic, palus2007directionality}
\begin{eqnarray}
\dot{x_{i}} &&= - y_i - z_i + S\sum_{j \neq i}^{N} A_{i j} \cdot\left(x_{j}-x_{i}\right) \nonumber\\ 
\dot{y_{i}} &&= x_{i}+ay_{i} \\ 
\dot{z_{i}} &&=-b + z_{i}(x_{i}-c) \nonumber
\label{eq:Lcon}
\end{eqnarray}
where $a=0.25$, $b=0.2$, $c=10$, $S=0.002$.
$\mathbf{A}=(A_{ij})$ is its adjacency matrix, with a connection density of 25\% throughout this work.
After simulating the Rössler network, we binarize $x_i(t)$ to obtain binary time series by setting up a binarization threshold $x^{\mathrm{th}}=8$ and a binning time step $\Delta t=3$ ms, similarly as those for Lorenz networks. In this work, time series of $x$-variable with length $2.7$ hours ($10^7\,\mathrm{ms}$) are used to reconstruct Rössler networks, with PDIF parameters set to $k=5$, $l=5$, $\tau=0$ ms (see Figs~\ref{fig:general_nets} and \ref{fig5}).

\subsubsection{Recurrent neural network}
The dynamics of the $i$-th node of a noise-driven
random recurrent neural network is described by
\begin{equation}
\tau_m \dot{x}_i = -x_i + S\sum_{j}A_{ij} \tanh(x_{j}) + \xi_i,
\end{equation}
where $\tau_m=20$ ms is the time constant, and $S=0.03$ is the coupling strength. 
$\mathbf{A}=(A_{ij})$ is its adjacency matrix, with a connection density of 25\%, and $\xi_i$ is the dimensionless independent Gaussian white noise for each node, independently sampled from $\mathcal{N}(0,0.01)$.
After simulating the network, we binarize $x_i(t)$ to obtain binary time series $T_i(k)$ by choosing a binarization threshold $x^\mathrm{th}=0.2$ 
and a binning time step $\Delta t=5$ ms, similarly as those for Lorenz networks.
In this work, time series of $x_i(t)$ with length $27$ hours ($10^8\,\mathrm{ms}$) are used to reconstruct networks, with PDIF parameters set to $k=4$, $l=1$, and $\tau=5$ ms (see Figs~\ref{fig:general_nets} and \ref{fig5}).

\subsection{Other compared methods}

\subsubsection{Symbolic transfer entropy (STE)}

STE is an extension of conventional TE \cite{staniek2008symbolic, martini2011inferring}.
Rather than operating on the continuous-valued activity time series,
STE first transforms the historical state vectors $y^{(k)}_n$ and $x^{(l)}_n$ 
into symbolic sequences through a preprocessing step, and then computes the TE
based on the resulting symbolized time series.
In contrast to PDIF's binarization, STE incorporates the idea of delayed embedding to capture temporal patterns by encoding them as discrete symbols,
which enhances its robustness to noise and non-stationarity compared with TE.
For instance, for an order-5 history sequence of node $x$, 
$x^{(5)}_n=\left[-0.75, 0.90, -0.19, 0.72, -0.44\right]$, the values are
ranked in ascending order by their indices, resulting in the symbol ``$15342$",
where each digit reflects the position of the original entity in the sorted sequences. 
Moreover, this symbolization significantly reduces the size of the state space 
from $5^5$ to $5!$, which helps mitigate the curse of dimensionality in estimating
the joint probabilities. 
For comparison with PDIF, we use history lengths $k=l=5$ and continuous-valued time
series with the identical data length as those used in PDIF's estimations for all benchmarks.

\subsubsection{GLMCC}

GLMCC is a network reconstruction method specifically designed for 
analyzing the spike train data \cite{kobayashi2019reconstructing}.
It combines cross-correlogram analysis with generalized linear model (GLM) methods to infer causal relationships between neurons.
To estimate the effective connection from node $X$ to node $Y$, GLMCC first computes the cross-correlogram 
$c_{X\to Y}(t)$ from the binarized time series (as generated in the PDIF processing pipelines).
It then fits a GLM of the form 
\begin{equation*}
 c(\tau) = \exp\left(a(\tau)+J_{YX}f(\tau)+J_{XY}f(-\tau)\right),
\end{equation*}
where $J_{YX}$ represents the effective coupling strength from node $X$ to $Y$, vice versa.
The function $f(\tau)$ models the spike-induced 
modulation of spike counts via an exponential kernel, while $a(t)$ captures background influences as slow-wave drives.
Typically, by fitting $J_{YX}$, GLMCC quantifies both strength and sign
(i.e., excitatory or inhibitory) of the effective interaction 
from $X$ to $Y$.
For our comparison, we adapt the Python implementation and hyperparameter settings reported in its original work \cite{kobayashi2019reconstructing}. For the HH networks, we compute GLMCC using data with the same length as used in PDIF cases.
For other networks, we used one-tenth of the full data length to reduce computational costs, which scale with the number of spike events in the binarized time series.

\subsubsection{Dynamical differential correlation (DDC)}

DDC estimates structural connectivity by integrating differential information with partial covariance analysis \cite{chen2022dynamical} .
The DDC-inferred connectivity matrix is computed as:
\begin{equation}
    W^\mathrm{DDC} = \mathrm{cov}(\dot{\mathbf{x}}, \mathbf{x})
    \mathrm{cov}(\mathbf{x}, \mathbf{x})^{-1}
\end{equation}
where $\mathrm{cov}(\cdot, \cdot)$ denotes the covariance matrix.
Conceptually, the first term captures the directional (driving-driven)
interactions through the relationship between a node’s derivative and the activity of other nodes, while the second term serves to mitigate the influence of common inputs.
DDC is particularly effective at identifying inhibitory causal interactions from relatively short recordings, but it relies on accurate estimation of time derivatives, which can make it sensitive to noise.
For a fair comparison with other methods evaluated in this study,
we used the same data length for DDC-based reconstruction across all networks as that used in PDIF.

\subsubsection{Convergent cross mapping (CCM) and its variations}

CCM infers causal relationships based on the theory of state-space reconstruction \cite{sugihara2012detecting,ye2015distinguishing, monster2017causal}.
Given discrete time series $\{x_n\}$ and $\{y_n\}$ with sampling
time step $\Delta t$ from node $X$ and $Y$, respectively,
we define the order-$l$ delay coordinate (DC) vectors as follows:
\begin{eqnarray*}
\mathbf{X}(n):=\left[x_{n},x_{n-1},...,x_{n-l+1}\right]^\top,\\
\mathbf{Y}(n):=\left[y_{n},y_{n-1},...,y_{n-l+1}\right]^\top.
\end{eqnarray*}
These vectors form delay-embedding manifolds $\mathbf{M}_X:=\{\mathbf{X}(t)|t\in[l-1,L]\}$ and
$\mathbf{M}_Y:=\{\mathbf{Y}(t)|t\in[l-1,L]\}$ in the state space.
If $X$ causally drives $Y$, the information about $\mathbf{M}_X$ should be reflected
in $\mathbf{M}_Y$, enabling the prediction of $\mathbf{M}_X$ based on the 
local geometry of $\mathbf{M}_Y$.
Practically, for the $n$-th delayed embedding state of $Y$, $\mathbf{Y}(n)$, we identify
its $l+1$ nearest neighbors in the state space, denoted as $\mathbf{Y}(n_1),\cdots,\mathbf{Y}(n_{l+1})$, with the discrete time index from $n_1$ to $n_{l+1}$.
Then, we predict $x_n$, the $n$-th state of $X$, using the information of
$\mathbf{Y}(n)$'s neighbors:
\begin{equation}
\hat{x}_n=\sum_{i=1}^{l+1} w_i x_{n_i},
\label{eq:CCM1}
\end{equation}
where $n_i$ refers to the time index of $i$-th nearest neighbor of $\mathbf{Y}(n)$ in the state space.
The weight $w_i$ is defined as
\begin{equation}
w_i=u_i / \sum_{j=1}^{l+1} u_j, \quad
u_i=\exp \left(-\frac{\left\|\mathbf{Y}(t)-\mathbf{Y}\left(t_i\right)\right\|}{\left\|\mathbf{Y}(t)-\mathbf{Y}\left(t_1\right)\right\|}\right).
\label{eq:CCM2}
\end{equation}
where $\lVert\cdot\rVert$ represents the Euclidean distance in the state space.
Intuitively, if $X$ causally influences $Y$, clusters of states on $\mathbf{M}_Y$ should map onto 
corresponding clusters on $\mathbf{M}_X$.
Therefore, $x_n$ can be predicted using the information from the
local structure of clusters on $\mathbf{M}_Y$.
As the data length $L$ increases, the prediction $\hat{x}_n$ is expected to converge to 
the true $x_n$.
In our implementation, we quantify the effective connection from $X$ to $Y$ using the Pearson correlation coefficient between $\{x_n\}$ and $\{\hat{x}_n\}$.

In addition, to address the sensitivity of CCM to noise, we also evaluate two 
recent variants of CCM:
Frequency-Domain CCM (FDCCM) \cite{avvaru2023effective} and Symbolic CCM
(SCCM) \cite{ge2023symbolic}.
FDCCM operates on DC vectors constructed in the frequency-domain, while SCCM replaces
the original DC vectors with symbolic permutation, similar to the symbolization step in STE.

Due to the high computational cost of the $k$-nearest neighbor search in CCM-based methods, we used one-tenth of the full time series length compared with PDIF reconstructions for all CCM, FDCCM, and SCCM reconstructions shown in Fig.~\ref{fig5}, to balance computational efficiency and performance.

\subsection{Derivation for the relation between PDIF and coupling strength}

For a two-node system with connection $X\to Y$,
the PDIF from $X$ to $Y$ is defined by Eq. (\ref{eq:TE}).
We first map the binary state of $y_{n}^{(k)}$ and $x_{n-\tau}^{(l)}$ to decimal numbers.
For an individual realization of $y_n^{(k)} = (y_n,y_{n-1},\cdots,y_{n-k+1})$, 
its decimal number representation is
\begin{equation*}
\sum_{i=1}^{k}y_{n+1-i}\cdot 2^{k-i}
\end{equation*}
We give the following notations
\begin{eqnarray*}
&&p(a,b)=p(y_{n}^{(k)}=a,x_{n-\tau}^{(l)}=b), \\
&&p(a)=p(y_{n}^{(k)}=a),\\
&&p_{a,b}=p(y_{n+1}=1|y_{n}^{(k)}=a,x_{n-\tau}^{(l)}=b), \\
&&\Delta p_{a,b}=p_{a,b}-p_{a,0},
% \label{eq:def_dp}
\end{eqnarray*}
where $a$ and $b$ are decimal numbers representing $y_{n}^{(k)}$ and $x_{n-\tau}^{(l)}$, respectively.
Then, we rewrite Eq. (\ref{eq:TE}) in the form of $p(a,b)$, $p_{a,b}$ and $\Delta p_{a,b}$ as
\begin{widetext}
\begin{eqnarray}
I_{X \to Y}&&=\sum_{a, b} \left[\underbrace{p_{a, b} p(a, b) \cdot \log \left(\frac{p_{a, b}}{\frac{\sum_b p_{a, b} \cdot p(a, b)}{p(a)}}\right)}_{\text{given } y_{n+1}=1}+\underbrace{\left(1-p_{a, b}\right) p(a, b) \cdot \log \left(\frac{1-p_{a, b}}{\frac{\sum_b\left(1-p_{a, b}\right) p(a, b)}{p(a)}}\right)}_{\text{given } y_{n+1}=0}\right]\nonumber\\
&&=-\sum_{a, b} \left[p_{a, b} p(a, b) \cdot \log \left(1+\frac{-\Delta p_{a,b}p(a)+\sum_b \Delta p_{a, b} {p(a,b)}}{p_{a, b}\cdot p(a)}\right)\right.\nonumber\\
&&\qquad\qquad\left.+\left(1-p_{a, b}\right) p(a, b) \cdot \log\left(1+\frac{\Delta p_{a,b}p(a)-\sum_b \Delta p_{a, b} {p(a,b)}}{(1-p_{a, b})\cdot p(a)}\right)\right].
\label{eq:TE_in_Pab}
\end{eqnarray}
\end{widetext}
Since $\Delta p_{a,b}$ represents the $X$'s activity induced change of state probability of $y_{n+1}=1$,
and the change is proportional to the synaptic strength which is rather weak in neurophysiological regime (numerically shown in the inset of Fig. \ref{fig:HH3}(e)),
$\Delta p_{a,b}$ should be a small quantity, and Eq. (\ref{eq:TE_in_Pab}) can be expanded with respect to $\Delta p_{a,b}$,
\begin{widetext}
\begin{eqnarray}
I_{X \to Y}&&=-\sum_{a, b} \left[p_{a, b} p(a, b) \left(\frac{-\Delta p_{a,b}p(a)+\sum_b \Delta p_{a, b} {p(a,b)}}{p_{a, b}\cdot p(a)}-\frac{\left(-\Delta p_{a,b}p(a)+\sum_b \Delta p_{a, b} {p(a,b)}\right)^2}{2p_{a, b}^2\cdot p(a)^2}\right)\right.\nonumber\\
&&\qquad\left.+\left(1-p_{a, b}\right) p(a, b) \left(\frac{\Delta p_{a,b}p(a)-\sum_b \Delta p_{a, b} {p(a,b)}}{(1-p_{a, b})\cdot p(a)}-\frac{(\Delta p_{a,b}p(a)-\sum_b \Delta p_{a, b} {p(a,b)})^2}{2(1-p_{a, b})^2\cdot p(a)^2}\right)+o(\Delta p_{a,b}^2)\right]\nonumber\\
&&=\frac{1}{2}\left[\sum_{a,b}\frac{p(a,b)}{p_{a,b}(1-p_{a,b})}\left(\Delta p_{a,b}^2-2\frac{\Delta p_{a,b}\left(\sum_b\Delta p_{a,b}p(a,b)\right)}{p(a)}+\frac{\left(\sum_b\Delta p_{a,b}p(a,b)\right)^2}{p(a)^2}\right)+o(\Delta p_{a,b}^2)\right]\nonumber\\
&&=\frac{1}{2}\sum_a\frac{1}{p_{a,0}-p_{a,0}^2}\sum_b \left[p(a,b)\left(\Delta p_{a,b}^2-2\frac{\Delta p_{a,b}\left(\sum_b\Delta p_{a,b}p(a,b)\right)}{p(a)}+\frac{\left(\sum_b\Delta p_{a,b}p(a,b)\right)^2}{p(a)^2}\right)+o(\Delta p_{a,b}^2)\right]\nonumber\\
&&=\frac{1}{2}\sum_a\frac{1}{p_{a,0}-p_{a,0}^2}\left[\sum_b p(a,b)\Delta p_{a,b}^2-2\frac{\left(\sum_b\Delta p_{a,b}p(a,b)\right)^2}{p(a)}+p(a)\frac{\left(\sum_b\Delta p_{a,b}p(a,b)\right)^2}{p(a)^2}+o(\Delta p_{a,b}^2)\right]\nonumber\\
&&= \frac{1}{2}\sum_{a}\frac{1}{p_{a,0}-p_{a,0}^{2}}\left(\sum_{b}p(a,b)\Delta p_{a,b}^{2}-\frac{\left(\sum_{b}p(a,b)\Delta p_{a,b})\right)^{2}}{p(a)}\right)+o\left(\Delta p_{a,b}^{2}\right).
\label{eq:TE expression1}
\end{eqnarray}
\end{widetext}
So far, according to Eq. (\ref{eq:TE expression1}), we identify the quadratic relationship between $I_{X\to Y}$ and $\Delta p_{a,b}$. 

Next, we demonstrate that $\Delta p_{a,b}$ in Eq. (\ref{eq:TE expression1}) further depends on the structural coupling strength $S$.
Without loss of generality, $\Delta p_{a,b}$ can be represented as a function of $S$, {i.e.}, $\Delta p_{a,b}=f(S)$. Expanding $\Delta p_{a,b}$ in terms of $S$ yields 
\begin{equation}
    \begin{aligned}
        \Delta p_{a,b} &= f(0)+f'(0)S+\frac{1}{2}f''(0)S^2+o(S^2) \\
        &=f'(0)S+\frac{1}{2}f''(0)S^2+o(S^2).
    \end{aligned}
    \label{eq:Pab_vs_S}
\end{equation}
The last equality in Eq. (\ref{eq:Pab_vs_S}) holds
because $\Delta p_{a,b} = 0$ when $S=0$.
Therefore, in the weakly coupled case (synaptic strength in neurophysiological regime), $\Delta p_{a,b}$ is proportional to $S$ at leading order. Consequently, combining Eq. (\ref{eq:TE expression1}) and (\ref{eq:Pab_vs_S}), we show that PDIF is proportional to $S^2$, as numerically
verified in Fig. \ref{fig:HH3}(e).

\subsection{Derivation for the relation between PDIF values of connected and unconnected pairs}
We demonstrate that PDIF can accurately 
identify directly connected pairs from others.
For the three-neuron HH network with chain motif in Fig. \ref{fig:HH3}(a), we show that the PDIF value $I_{X\to Z}$ is orders of magnitude smaller than $I_{X\to Y}$
(or $I_{Y\to Z}$).
Assume that the coupling strength from $X$ to $Y$ and from $Y$ to $Z$ are $S_1$ and $S_2$, respectively,
the $X$'s spike induced change of firing probability of $Z$, $\Delta p^{X\to Z}_{a,b}$ should depend on both $S_1$ and $S_2$, i.e., $\Delta p^{X\to Z}_{a,b}=g(S_1, S_2)$. 
Treating $S_1$ and $S_2$ as small quantities, 
$g(S_1, S_2)$ can be Taylor expanded as
\begin{eqnarray*}
\Delta p_{a,b}^{X\to Z} =&& g(0,0)+\left.\frac{\partial g}{\partial S_1}\right|_{(0,0)}S_1+\left.\frac{\partial g}{\partial S_2}\right|_{(0,0)}S_2\nonumber\\
&&+\frac{1}{2}\left.\frac{\partial^2 g}{\partial S_1^2}\right|_{(0,0)}S_1^2+\frac{1}{2}\left.\frac{\partial^2 g}{\partial S_2^2}\right|_{(0,0)}S_2^2\nonumber\\
&&+\frac{1}{2}\left.\frac{\partial^2 g}{\partial S_1\partial S_2}\right|_{(0,0)}S_1S_2+o(S_1S_2).\\
\end{eqnarray*}
Since $\Delta p^{X\to Z}_{a,b}$ is nonzero if and only if both $S_1$ and $S_2$ are nonzero, its leading term is the bilinear product of $S_1$ and $S_2$, i.e.,
\begin{equation}
    \Delta p_{a,b}^{X\to Z} = \frac{1}{2}\left.\frac{\partial^2 g}{\partial S_1\partial S_2}\right|_{(0,0)}S_1S_2+o(S_1S_2).
    \label{eq:Pab_vs_S1S2}
\end{equation}
Since $\Delta p^{X\to Y}\propto O(S_1)$ and $\Delta p^{Y\to Z}\propto O(S_2)$, we have
\begin{equation*}
    \Delta p^{X\to Z}=O(\Delta p^{X\to Y}\cdot \Delta p^{Y\to Z})
\end{equation*}
by retaining the dominant term in the expansion.
In networks with coupling strengths in the neurophysiological regime, both $\Delta p^{X\to Y}_{a,b}$ and $\Delta p^{Y\to Z}_{a,b}$
for direct connections are relatively small.
As shown in the inset of Fig. \ref{fig:HH3}(e), these values are typically less than 0.01 for the HH network. Consequently, the indirect increment $\Delta p^{X\to Z}_{a,b}$ is significantly smaller than $\Delta p^{X\to Y}_{a,b}$, a result numerically confirmed in Fig. \ref{fig:HH3}(i).  
Combined with the derived relation in Eq. (\ref{eq:TE expression1}), $I_{X\to Z}$ is orders of magnitude smaller than $I_{X\to Y}$. This ensures robust differentiation between direct and indirect interactions, preventing false positive inferences. The orders of magnitude difference between $I_{X\to Z}$ and $I_{X\to Y}$ is further validated in the inset of Fig. \ref{fig:HH3}(i).
Similarly, for the confounder motif (Fig. \ref{fig:HH3-confounder}(a)), we can also derive that $\Delta p^{Y\to Z}=O(\Delta p^{X\to Y}\cdot \Delta p^{X\to Z})$, which is verified numerically in Fig. \ref{fig:HH3}(j) (inset: corresponding PDIF relations). 

In summary, unlike information-theoretic measures (such as TE and conditional TE),
PDIF based on binary time series effectively
distinguishes the effective connectivity arising from
direct physical interactions  versus indirect ones. 
Importantly, it achieves this without relying on information about the third-party node, ensuring robust and effective network reconstruction.

\begin{acknowledgments}
This work is supported by National Key R\&D Program of China 2023YFF1204200, Science and Technology Commission of Shanghai Municipality with Grant No. 24JS2810400 (D.Z., and S.Li); National Natural Science Foundation of China with Grant Nos. 92570202 and 12225109 (D.Z.); National Natural Science Foundation of China Grants 12271361 and 12250710674 (S.Li); and the Student Innovation Center at Shanghai Jiao Tong University (K.C., Z.T., Y.C., S.L., S.Li, and D.Z.).
\end{acknowledgments}

\bibliography{B-PTD-TE-nc}% Produces the bibliography via BibTeX.

\renewcommand{\thefigure}{S\arabic{figure}}
\renewcommand{\thetable}{S\arabic{table}}
\renewcommand{\theequation}{S\arabic{equation}}
\setcounter{figure}{0}
\setcounter{table}{0}
\setcounter{equation}{0}

\clearpage
\onecolumngrid
\appendix
\section{Supplementary Figures}
\begin{figure*}[!htb]
    \centering
    \includegraphics[width=1\linewidth]{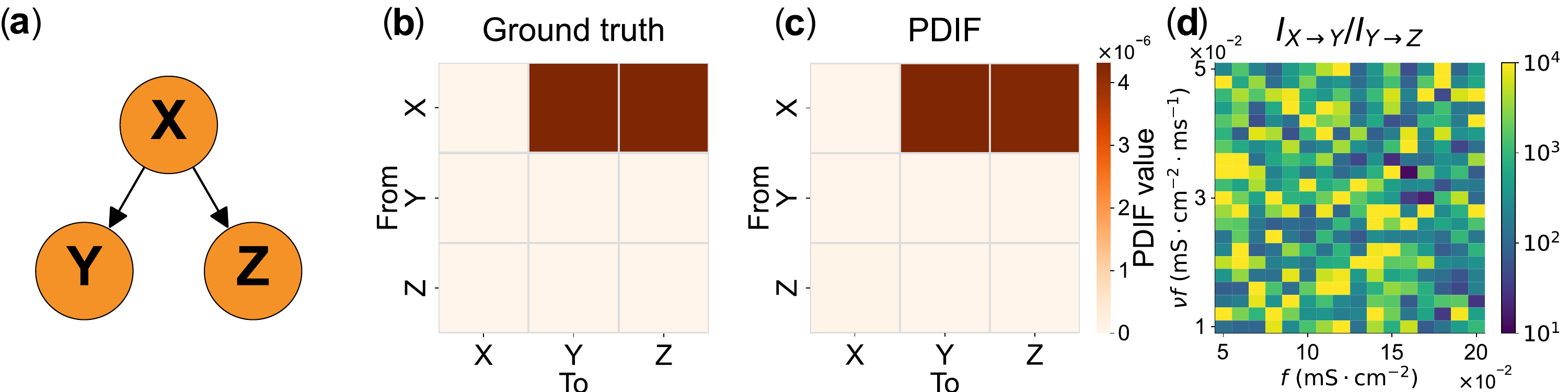}
    \caption{\textbf{PDIF reconstruction performance for three-neuron HH network with confounder motifs.}
    (\textbf{a}) The graph of a three-neuron network with confounder motif.
    (\textbf{b}) The adjacency matrix for the network in (\textbf{a}).
    (\textbf{c}) The matrix of inferred PDIF values for the network in (\textbf{a}). 
    (\textbf{d}) The ratio of PDIF values between connected and unconnected pairs, $I_{X\to Y}/I_{Y\to Z}$. 
	The minimum ratio in (\textbf{d}) is greater than 10,
    across a broad dynamical regime with firing rates from $2$ to $50$ Hz.
    In these cases, $2.7$ hours neuronal recordings are used for PDIF estimation. 
    PDIF parameters are chosen as $k=1$, $l=1$, and $\tau=3$ ms, following the
    pipeline in Fig. \ref{fig:recon-pipeline} in the main text.
    }\label{fig:HH3-confounder}
\end{figure*}

\begin{figure*}[!htb]
    \includegraphics[width=\textwidth]{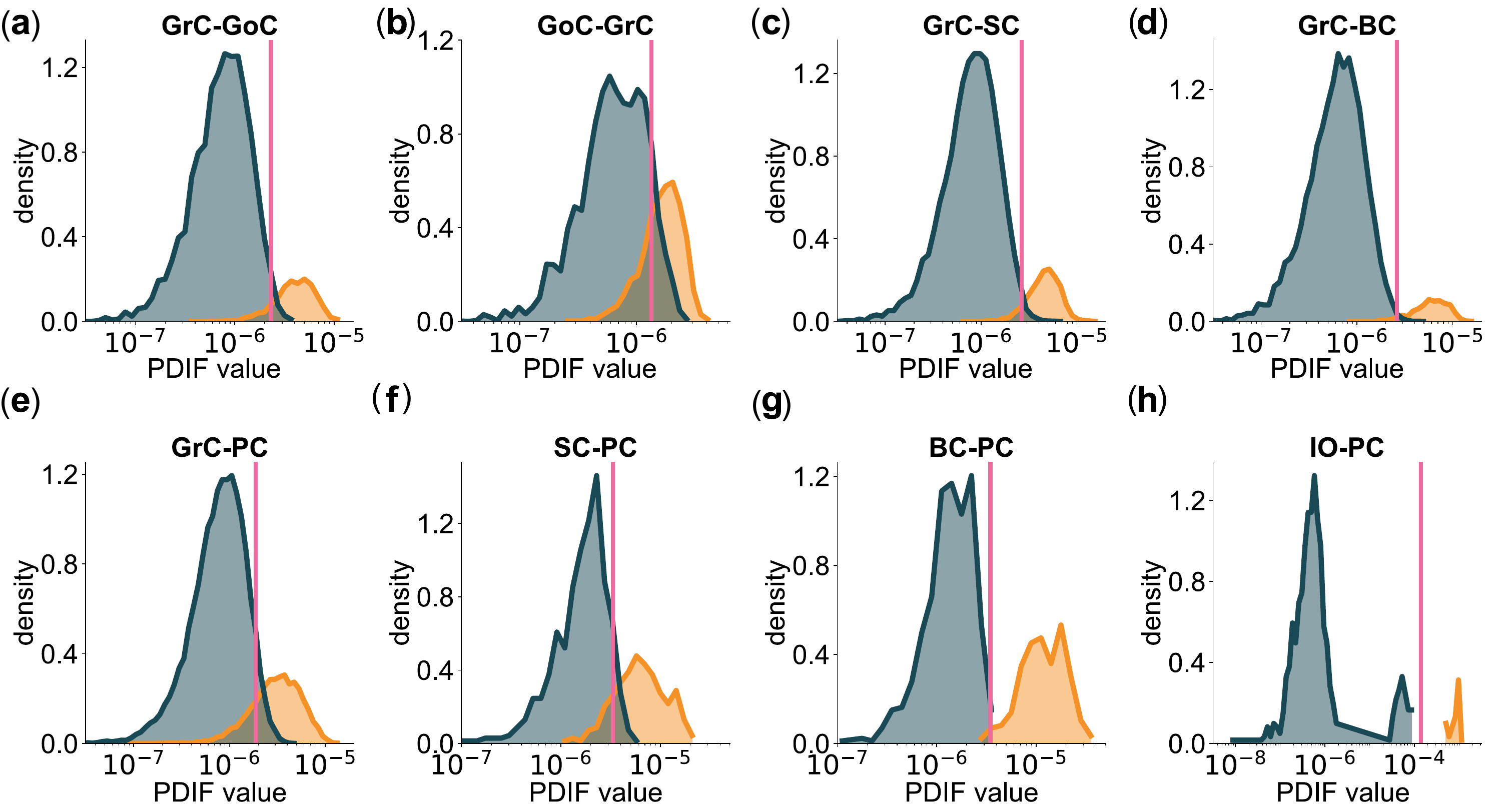}
    \caption{
    \textbf{PDIF reconstruction performance applied to the cerebellar network.}
    PDIF parameters are set to $k=1$, $l=5$, with $\Delta t = 0.1$ ms.
    The optimal reconstruction delay parameters ($\tau$) and the corresponding reconstruction accuracies 
    are listed in Tab.~\ref{tab:projection-params}.
    The orange and dark green histograms represent the distribution of PDIF values for the connected and unconnected neuronal pairs, respectively.
    The pink vertical lines indicate the optimal reconstruction thresholds determined via GMM (see Methods).
    }
    \label{fig:cerebellar_full_TE}
\end{figure*}

\begin{figure*}[!htb]
	\includegraphics[width=\textwidth]{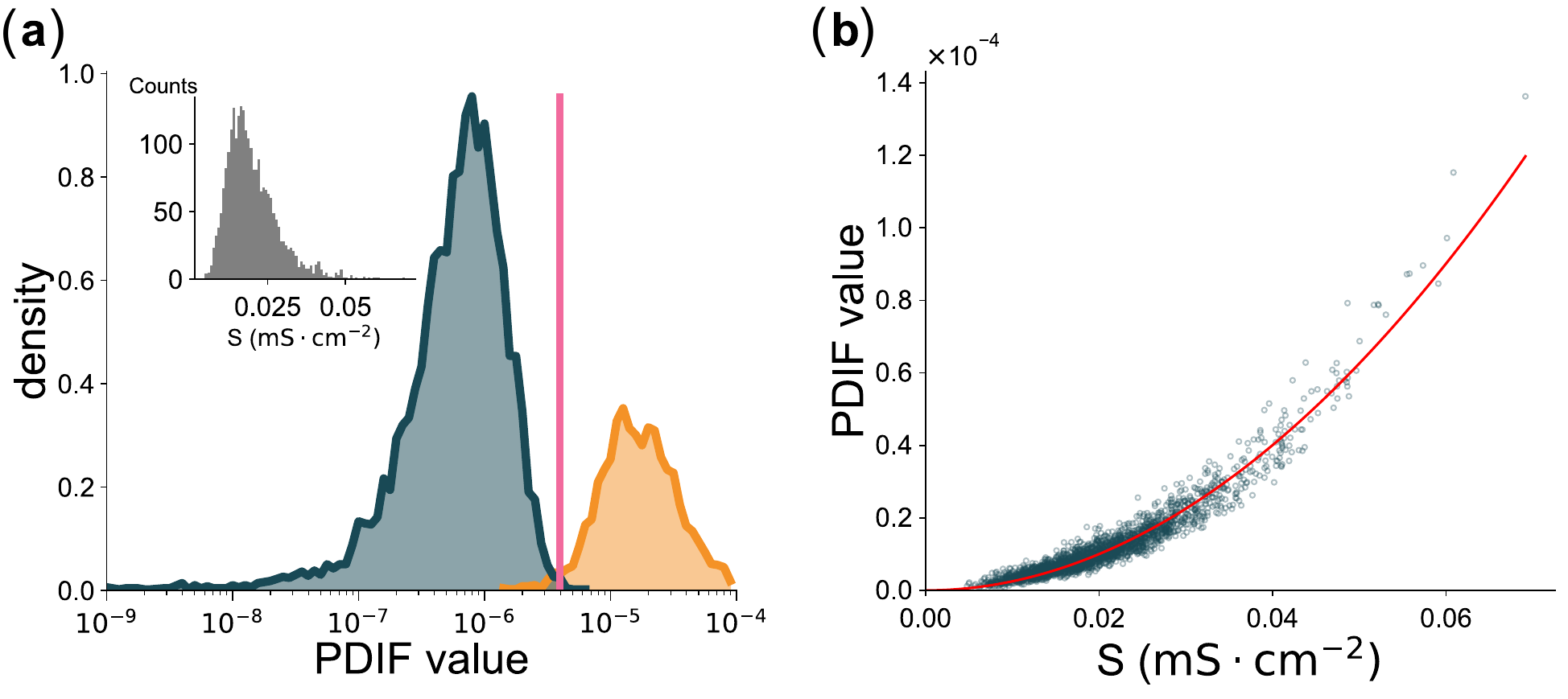}
	\caption{
        \textbf{PDIF reconstruction of an HH network with log-normally distributed structural connectivity.} 
        (\textbf{a}). The distribution of PDIF values.
        PDIF parameters are set to $k=l=1$, $\tau=3$ ms, the same as in Fig. \ref{fig:HH100}(a) in the main text. Orange and dark green corresponds to PDIF values of connected and unconnected pairs, respectively.
        The pink line indicates the optimal reconstruction threshold, achieving 99.26\% reconstruction accuracy.
        Inset: Histogram of log-normally distributed coupling strengths $S$ in the network,
        and remaining parameters for the HH network simulation match those in
        100-neuron HH networks in the main text. 
        (\textbf{b}). PDIF values versus coupling strength $S$. The quadratic relation between PDIF and $S$ also valid for
        networks with heterogeneous structural connectivity. 
		The red solid line is a quadratic fit with $R^2=0.94$.
		}
		\label{fig:log-normal}
\end{figure*}

\begin{figure*}[!htb]
    \includegraphics[width=\textwidth]{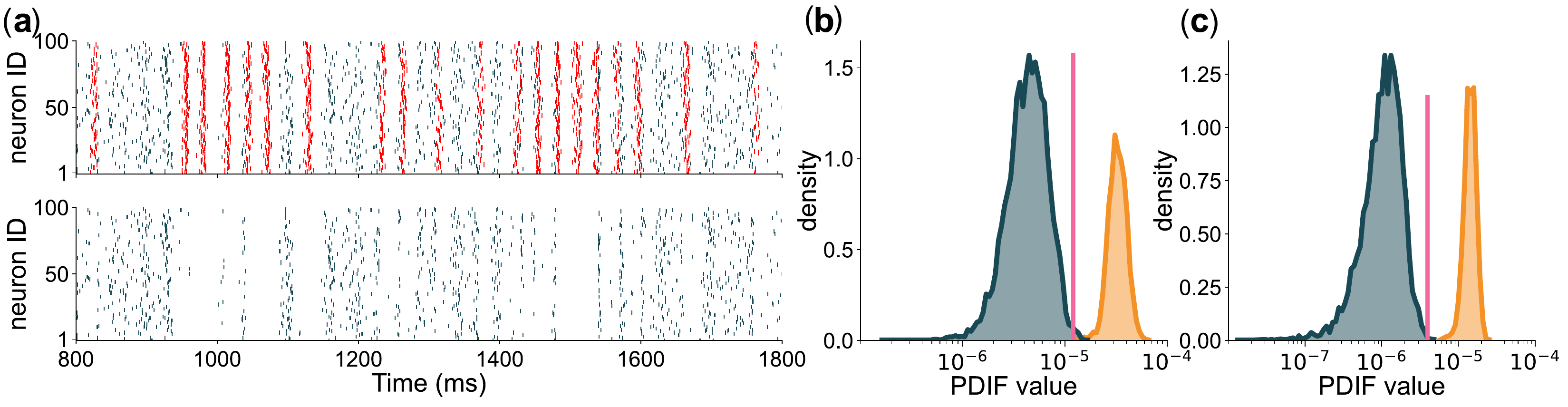}
    \caption{
        \textbf{Reconstructions for HH networks in near synchronous regime.}
        (\textbf{a}) The raster plot of an HH network in quasi-synchronous regime with strong recurrent
        coupling ($S=0.028\,\mathrm{mS}\cdot\mathrm{cm}^{-2}$) (top) before and (bottom) after down-sampling.
        All spikes are uniformly binned, with 10 ms bin size, into time series $r(t)$ to quantify the synchronization level, where $r(k)$ is the total number of spikes in the network in $k$-th time bin.
        We denote $k$-th bin as the synchronized bins if $r(k) \ge \langle r\rangle +0.5\,\sigma(r)$, where $\langle r\rangle$ and $\sigma(r)$ are the mean and standard deviation of spike counts within single bin over time.
        To improve the inference performance for networks in quasi-synchronous state, the down-sampling operation removes spikes within synchronized time bins, indicated by red bars in the top subfigure.
        (\textbf{b}) The distribution of PDIF values for the same HH network in (\textbf{a})
        (\textbf{c}) The distribution of PDIF values for another quasi-synchronous HH network, induced by moderate correlated Poisson inputs (35\% input correlation with strength $f=0.2 \,\mathrm{mS}\cdot\mathrm{cm}^{-2}$ and frequency $\nu=0.05\,\mathrm{ms}^{-1}$), after similar down-sampling preprocessing.
        PDIF parameters are chosen
        as $k=l=1$, $\tau=3$ ms, $\Delta t=0.5$ ms in (\textbf{b})-(\textbf{c}).
        The adjacency matrices for structural connectivity are the same as in Fig. \ref{fig:HH100} in the main text.
        Orange and dark green curves are corresponding to PDIF values of connected and unconnected pairs, respectively. Pink lines indicate the optimal thresholds,
        resulting 99.53\% and 99.93\% reconstruction accuracy for (\textbf{b}) and (\textbf{c}), respectively.
    }
    \label{fig:downsample}
\end{figure*}

\begin{figure*}[!htb]
    \centering
    \includegraphics[width=1.0\linewidth]{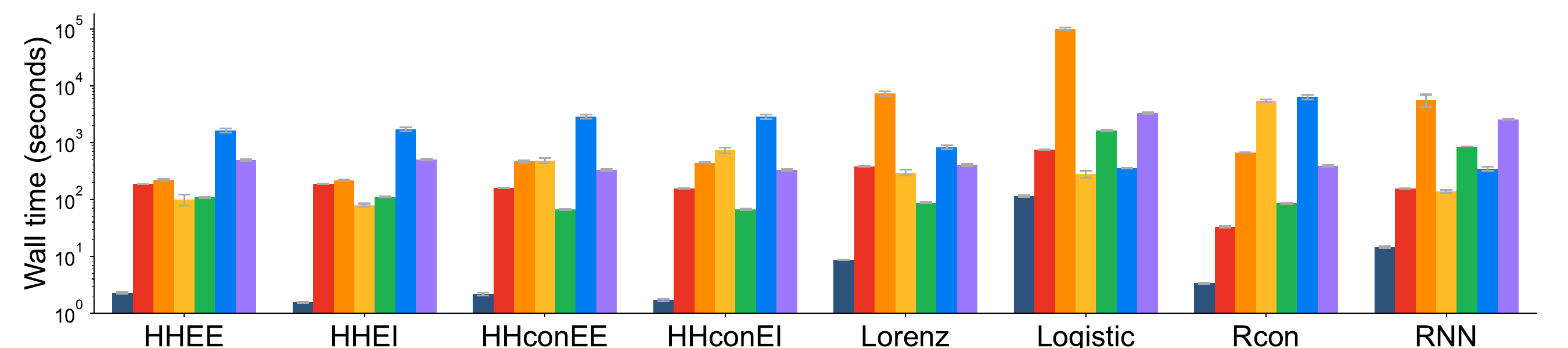}
    \caption{
    \textbf{The average wall time required for each reconstruction method to reconstruct various networks.}
    The error bars indicate the standard error of mean among all reconstructions among different noise levels and trials.
    The reconstruction for each trial is conducted exclusively on a computing node with  
    two Intel Xeon Cascade Lake 6248 chips (2.5GHz, 40 cores in total) and 192 GB DDR4 RAMs.
    Color codes indicating reconstruction methods match with those in Fig.~\ref{fig5}.
    }
    \label{fig:wall-time}
\end{figure*}

\begin{figure}[!htb]
    \centering
    \includegraphics[width=0.6\linewidth]{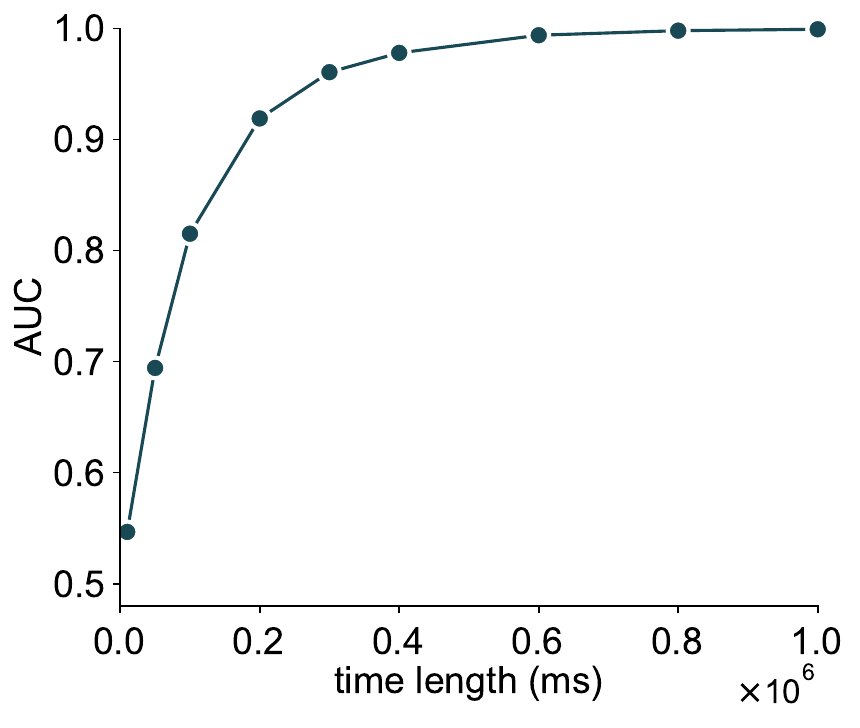}
    \caption{\textbf{Relation between PDIF performance and the data length used for reconstruction.}
    The analysis is performed using the same 100-neuron HH network as those in Fig.~\ref{fig:HH100}(a).}
    \label{fig:HH100_TE_vs_length}
\end{figure}

\begin{figure}[!htb]
    \centering
    \includegraphics[width=0.6\linewidth]{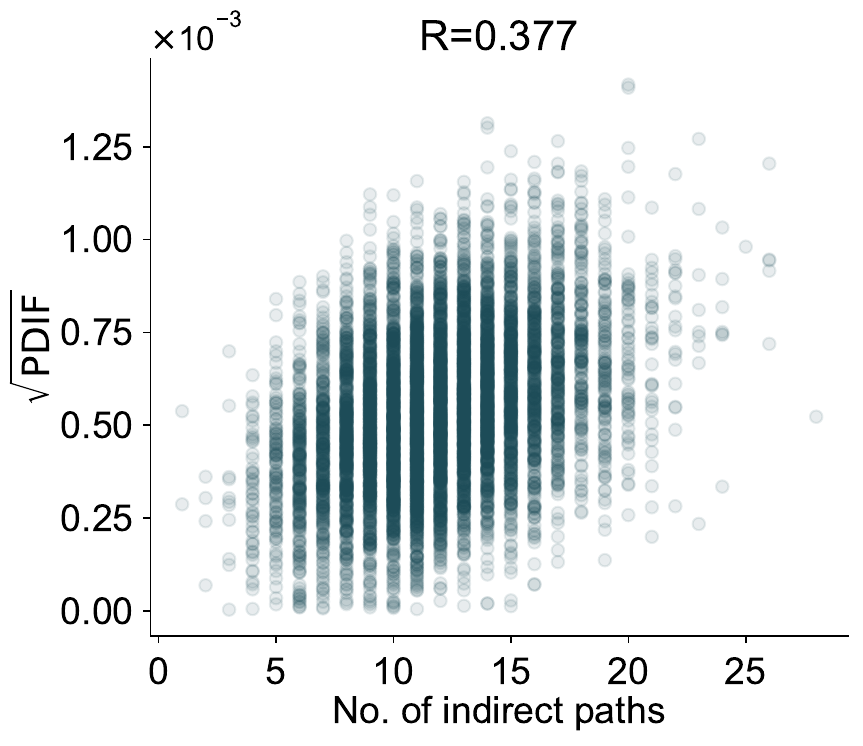}
    \caption{\textbf{Relation between PDIF value and the number of indirect interaction paths for indirectly coupled neuron pair in HH network.}
    The same 100-neuron HH network as those in 
    Fig.~\ref{fig:HH100}(a) of the main text
    is analyzed, and both chain and confounder motifs are included for analysis.
    The correlation between $\sqrt{\text{PDIF}}$ and number of indirect paths is $0.377$.
    }
    \label{fig:HH100_indirect}
\end{figure}

\newpage
\FloatBarrier
\section{Supplementary Tables}
\begin{table}[h]
\centering
% \captionsetup{width=\textwidth}
\caption{\textbf{Neuronal populations in the cerebellar circuit network.}
Neuronal populations and background noise (BG) inputs with their respective neuron numbers, mean firing rates, and model types.
% and input sources with corresponding numbers and firing rates.
}
\begin{tabular}{c c c c}
\toprule
Unit & Num & Fr (Hz) & Type\\
\hline
GrC & 1024 & 2.12   & multi-compartment\\
GoC & 16   & 18.62  & multi-compartment\\
BC  & 24   & 24.80  & multi-compartment \\
SC  & 24   & 19.12  & multi-compartment\\
PC  & 36   & 49.81  & multi-compartment\\
DCN & 8    & 14.45  & multi-compartment\\
IO  & 24   & 2.96   & single-compartment\\
MF  & 1024 & 12.50  & Poisson generator\\
BG  & 48   & 0.99   & Poisson generator\\
\botrule
\end{tabular}
\label{tab:cerebellar-pop}
\end{table}

\begin{table*}[ht]
\centering
% \captionsetup{width=\textwidth}
\caption{
\textbf{Projection and synaptic parameters in the cerebellar circuit network.}
Projections are characterized by the projection type (Proj), in-degree (In-deg), and targeted synaptic location (Syn loc).
Synaptic dynamics parameters are characterized by transmission delay (Delay), time constants ($\tau_1$, $\tau_2$), reversal potential ($E_{\mathrm{rev}}$), synaptic weight, and the resulting postsynaptic potential (PSP).
For synaptic locations defined as intervals, the range indicates the normalized path length along the target compartment where synapses are randomly distributed. Corresponding PSP values are reported as ranges to reflect the variability arising from these spatial distributions. All projection pairs utilize a single synapse per connection, with the exception of IO--PC (20 synapses) and IO--DCN (4 synapses), denoted as $\textit{weight} \times \textit{num}$ in the Weight column.
}
\begin{tabular}{c c c c c c c c}
\toprule
Proj & In-deg & Syn loc & Delay(ms) & $\tau_1,\tau_2$ (ms) & $E_{\mathrm{rev}}$ (mV) & Weight & PSP (mV) \\
\hline
BG-IO    & 1   & Soma        & 2 & 0.5, 3 & 0       & 0.0002          & 3.6             \\
MF-DCN   & 80  & Distdend    & 2 & 0.5, 3 & 0       & 0.001           & (1.06, 1.17)    \\
MF-GrC   & 4   & Dend        & 2 & 0.5, 3 & 0       & 0.0012          & 2.5             \\
MF-GoC   & 100 & Dend        & 2 & 0.5, 3 & 0       & 0.0001          & (0.08, 0.13)    \\
GrC-GoC  & 100 & Dend (0, 0.5) & 3 & 0.5, 3 & 0     & 0.002           & (1.3, 2.5)      \\
GoC-GrC  & 4   & Dend        & 2 & 0.8, 4 & -75     & 0.00014         & -1.1            \\
GrC-SC   & 100 & Dend        & 3 & 0.5, 3 & 0       & 0.00045         & (2.0, 2.4)      \\
GrC-BC   & 50  & Dend        & 3 & 0.5, 3 & 0       & 0.00035         & (1.7, 2.1)      \\
GrC-PC   & 200 & Dend (0.1, 0.8) & 3 & 0.5, 3 & 0   & 0.006           & (0.7, 1.3)      \\
SC-PC    & 7   & Dend (0.4, 0.9) & 2 & 0.8, 4 & -75 & 0.025           & (-0.8, -0.4)    \\
BC-PC    & 7   & Dend (0, 0.5) & 4 & 0.8, 4 & -75   & 0.02            & (-0.9, -0.5)    \\
PC-DCN   & 10  & Proxdend    & 3 & 0.8, 6 & -75     & 0.006           & (-1.84, -1.74)  \\
DCN-IO   & 1   & Soma        & 3 & 0.8, 6 & -75     & 0.0003          & -0.475          \\
IO-DCN   & 1   & Proxdend    & 2 & 0.5, 3 & 0       & 0.0025$\times$4 & (2.86, 2.89)    \\
IO-PC    & 1   & Dend (0, 0.5) & 2 & 0.5, 3 & 0     & 0.005$\times$20 & (0.94, 1.1)     \\
\botrule
\end{tabular}
\label{tab:projection-params}
\end{table*}

\begin{table}[htp]
% \captionsetup{width=\textwidth}
\centering
\caption{\textbf{PDIF reconstruction performance.}
Performance for different cross-population projections, showing accuracy (ACC) and area under the ROC curve (AUC), evaluated with optimal time delay ($\tau$).
}
\begin{tabular}{c c c c}
\toprule
Proj & $\tau$ & ACC & AUC \\
\hline
GrC-GoC & 8 & 97.30\% & 0.9852 \\
GoC-GrC & 9 & 85.67\% & 0.9093 \\
GrC-BC & 7 & 99.56\% & 0.9986 \\
GrC-SC & 7 & 98.38\% & 0.9949 \\
GrC-PC & 6 & 91.30\% & 0.9426 \\
BC-PC & 2 & 99.31\% & 0.9994 \\
SC-PC & 6 & 90.74\% & 0.9575 \\
PC-DCN & 8 & 95.49\% & 0.9944 \\
DCN-IO & 13 & 98.96\% & 0.9995 \\
IO-DCN & 4 & 100.00\% & 1.0000 \\
IO-PC & 3 & 100.00\% & 1.0000 \\
\botrule
\end{tabular}
\label{tab:cerebellar-perf}
\end{table}
% \nocite{*}

\end{document}